\documentclass{amsproc}
\usepackage{amsmath, amsthm, amssymb, slashed}
\usepackage{ifpdf}
\ifpdf
  \usepackage[pdftex]{graphicx}
  \usepackage{epstopdf}
\else
  \usepackage[dvips]{graphicx}
\fi
\usepackage{url}
\newtheorem{theorem}{Theorem}[section]

\theoremstyle{definition}

\theoremstyle{remark}
\newtheorem{remark}[theorem]{Remark}

\numberwithin{equation}{section}

\title{Geometric Langlands And The Equations Of Nahm And Bogomolny}

\author{Edward Witten}
\address{Theory Group, CERN, Geneva Switzerland.  On leave from
School of Natural Sciences, Institute for Advanced Study, Princeton
NJ 08540 USA. }
\email{witten@ias.edu}
\thanks{Theory Group, CERN, Geneva Switzerland.  On leave from School of Natural
Sciences, Institute for Advanced Study, Princeton NJ 08540.
Supported in part by NSF Grant Phy-0503584.}

\font\teneurm=eurm10 \font\seveneurm=eurm7 \font\fiveeurm=eurm5
\newfam\eurmfam
\textfont\eurmfam=\teneurm \scriptfont\eurmfam=\seveneurm
\scriptscriptfont\eurmfam=\fiveeurm

 \font\teneusm=eusm10 \font\seveneusm=eusm7 \font\fiveeusm=eusm5
\newfam\eusmfam
\textfont\eusmfam=\teneusm \scriptfont\eusmfam=\seveneusm
\scriptscriptfont\eusmfam=\fiveeusm
\def\eusm#1{{\fam\eusmfam\relax#1}}
\font\tencmmib=cmmib10 \skewchar\tencmmib='177
\font\sevencmmib=cmmib7 \skewchar\sevencmmib='177
\font\fivecmmib=cmmib5 \skewchar\fivecmmib='177
\newfam\cmmibfam
\textfont\cmmibfam=\tencmmib \scriptfont\cmmibfam=\sevencmmib
\scriptscriptfont\cmmibfam=\fivecmmib
\def\cmmib#1{{\fam\cmmibfam\relax#1}}

\def\tilde{\widetilde}
\def\R{{\Bbb{R}}}\def\Z{{\Bbb{Z}}}

\begin{document}
\def\C{\Bbb{C}}
\def\M{{\cal M}}
\def\Bbb{\mathbb}
\def\frak{\mathfrak}
\subjclass{}
\date{February, 2009}

\begin{abstract}
Geometric Langlands duality relates a representation of a simple Lie
group $G^\vee$ to the cohomology of a certain moduli space
associated with the dual group $G$.  In this correspondence, a
principal $SL_2$ subgroup of $G^\vee$ makes an unexpected
appearance.  Why this happens can be explained using gauge theory,
as we will see in this article, with the help of the equations of
Nahm and Bogomolny.  (Based on a lecture at Geometry and Physics:
Atiyah 80, Edinburgh, April 2009.)

\end{abstract}

\maketitle
\def\D{{\mathcal D}}
\def\Z{{\Bbb Z}}
\def\hat{\widehat}
\def\NN{\eusm N}

\section{Introduction}

This article is intended as an introduction to the gauge theory approach \cite{KW} to the geometric Langlands
correspondence.  But rather than a conventional overview, such as I have attempted elsewhere \cite{W1,W2},
here I will focus on explaining what we need to understand a very particular result, which I learned of from the expository article \cite{G}.  (Another standard reference on closely related matters is \cite{MV}.)

The present introduction will be devoted to describing the facts
that we wish to explain.  In  section \ref{gaugetheory}, gauge
theory -- in the form of new results about how duality acts on
boundary conditions \cite{GW1,GW2} -- will be brought to bear to
explain them. Finally, some technical details are reserved for
section \ref{details}.  In section \ref{reduction}, we also briefly
discuss compactification of the relevant gauge theory to three
dimensions, showing some novel features that appear to be relevant
to recent work \cite{BN}.  And in section \ref{genker}, we discuss
the universal kernel of geometric Langlands from a gauge theory
point of view.

\def\Hom{{\mathrm{Hom}}}

\subsection{The Dual Group}\label{dualg}
Let us start with a compact simple Lie group $G$ and its Langlands or Goddard-Nuyts-Olive dual group $G^\vee$.
(In gauge theory, we start with a compact gauge group $G$, but by the time we make contact with the usual statements
of geometric Langlands, $G$ is replaced by its complexification $G_\C$.)
If we write $T$
and $T^\vee$ for the respective maximal tori, then the basic relation between them is that
\begin{equation}\label{poly}\Hom(T^\vee,U(1))=\Hom(U(1),T),\end{equation}
and vice-versa.    Modulo some standard facts about simple Lie groups, this relation defines the correspondence
between $G$ and $G^\vee$.

Now let $R^\vee$ be an irreducible representation of $G^\vee$.  Its highest weight is a homomorphism $\rho^\vee:T^\vee\to
U(1)$.  Via (\ref{poly}), this corresponds to a homomorphism in the opposite direction $\rho:U(1)\to T$.

\def\CP{\Bbb{CP}}
\def\C{\Bbb{C}}
\def\L{{\mathcal L}}
\def\N{{\mathcal N}}
We can think of $U(1)\cong S^1$ as the equator in $S^2\cong\CP^1$.  With this understood, we can view $\rho:S^1\to G$
as a ``clutching function'' that defines a holomorphic $G_\C$ bundle  $E_\rho\to \CP^1$.   Every holomorphic $G_\C$ bundle over
$\CP^1$ arises this way, up to isomorphism, for a unique choice of $R^\vee$.  Thus, isomorphism classes of such bundles
correspond to isomorphism classes of irreducible representations of $G^\vee$.  In the language of Goddard, Nuyts, and Olive,
this is the correspondence between electric charge of $G^\vee$ and magnetic charge of $G$.

Actually, the homomorphism $\rho:U(1)
\to G$ can be complexified  to a homomorphism $\rho:\C^*\to G_\C$.  Here, we can view $\C^*$ as the complement in
$\CP^1$ of two points $p$ and $q$ (the north and south poles).   So the bundle $E_\rho$ is naturally made by gluing a trivial bundle over
$\CP^1\backslash p$ to a trivial bundle over $\CP^1\backslash q$.  In particular, $E_\rho$ is naturally trivial over the complement of the
point $p\in \CP^1$.  So  $E_\rho$ is ``a Hecke modification at $p$ of the trivial $G_\C$ bundle over $\CP^1$.''  By
definition, such a Hecke modification is simply a holomorphic $G_\C$ bundle $E\to \CP^1$ with a trivialization over the
complement of $p$.  $E$ is said to be of type $\rho$ if, forgetting the trivialization, it is equivalent holomorphically to $E_\rho$.

More generally, for any Riemann surface $C$, point $p\in C$, and holomorphic $G_\C$ bundle $E_0\to C$,
a Hecke modification of $E_0$ at $p$ is a holomorphic $G_\C$ bundle $E\to C$ with an isomorphism $\varphi:E\cong E_0$
away from $p$. As in \cite{G,MV}, loop groups and affine Grassmannians give a natural language for describing these notions,
and explaining in general what it means to say that a Hecke modification is of type $\rho$.  We will not need this language here.

\subsection{An Example}
Let us consider an example.  Suppose that $G^\vee=SU(N)$ for some $N$, and accordingly its complexification is
$G_\C^\vee=SL(N,\C)$.   Then $G=PSU(N)$ and $G_\C=PSL(N,\C)$.  We can think of a holomorphic $G_\C$ bundle
as a rank $N$ holomorphic vector bundle $V$, with an equivalence relation $V\cong V\otimes\L$, for any holomorphic
line bundle $\L$.  (The equivalence relation will not play an important role in what we are about to say.)  Let us take
the representation $R^\vee$ to be the obvious $N$-dimensional representation of $G^\vee_\C=SL(N,\C)$.
To this data, we should associate a rank $N$ holomorphic bundle $V\to\CP^1$ that is obtained by modifying the
trivial bundle $U=\C^N\times \CP^1\to \CP^1$ at a single point $p\in \CP^1$.  More precisely, we will get a family of
possible $V$'s -- the possible Hecke modifications of $U$ of the appropriate type.
To describe such a $V$, pick a one-dimensional complex subspace $S\subset \C^N$ and let $z$ be a local coordinate
near $p$.  And declare that a holomorphic
section $v$ of $V$ over an open set $\mathcal U\subset \CP^1$ is a  holomorphic section of $U$ over $\mathcal U\backslash p$
which near $p$ looks like
\begin{equation}\label{dolly} v = a +\frac{s}{z},\end{equation}
where  $a$ and $s$ are holomorphic at $z=0$ and $s(0)\in S$.

This gives a Hecke modification of $U$, since $V$ is naturally equivalent to $U$ away from $z=0$.  Clearly, the definition
of $V$ depends on $S$, so we have really constructed a family of possible $V$'s, parametrized by $\Bbb{CP}^{N-1}$.
This is the family of all possible Hecke modifications of the appropriate type.

There is an analog of this for any choice of representation $R^\vee$ of the dual group.  To such a representation,
we associate as before the clutching function $\rho:U(1)\to T\subset G$, leading to a holomorphic $G_\C$ bundle
$E_\rho\to \Bbb{CP}^1$.  Then we define $\N(\rho)$ to be the space of all possible\footnote{In one important respect,
our example was misleadingly simple.
In our example, every possible Hecke modification can be made using a clutching function associated with a homomorphism
$\widetilde\rho:\C^*\to G_\C$ (which is conjugate to the original homomorphism $\rho:\C^*\to T_\C\subset G_\C$).  Accordingly,
in our example, $\N(\rho)$ is a homogeneous space for an obvious action of $G_\C$.  In general, this
is only so if the representation  $R^\vee$ is ``minuscule,'' as it is in our example.} Hecke modifications at $p$ of the trivial bundle
over $\Bbb{CP}^1$ that are of type $\rho$.

\def\bar{\overline}
\def\bN{\bar{\N}}
The moduli space $\N(\rho)$ of possible Hecke modifications has a
natural compactification $\bN(\rho)$. In the description of
$\bN(\rho)$ via the three-dimensional Bogomolny equations, which we
come to in section \ref{gaugetheory}, the compactification involves
monopole bubbling,\footnote{This phenonomenon was investigated in
the 1980's in unpublished work by P. Kronheimer, and more recently
in \cite{KW,CD}.} which is analogous to instanton bubbling in four
dimensions. $\N(\rho)$ is known as a Schubert cell in the affine
Grassmannian, and $\bN(\rho)$ as a Schubert cycle in that
Grassmannian. $\bN(\rho)$ parametrizes a family of Hecke
modifications of the trivial bundle, but they are not all of type
$\rho$; the compactification is achieved by allowing Hecke
modifications dual to a representation of $G^\vee$ whose highest
weight is ``smaller'' than that of $R^\vee$.

\subsection{The Principal ${\cmmib{SL}_{\mathbf 2}}$}\label{principal}
Geometric Langlands duality associates the representation $R^\vee$ of the dual group to the cohomology of $\bN(\rho)$.
Let us see how this works in our example.

In the example, $R^\vee$ is the natural $N$-dimensional representation of $SL(N,\C)$, and $\N(\rho)$ (which  needs no compactification, as $R^\vee$ is minuscule) is $\CP^{N-1}$  Not coincidentally, the cohomology of
$\CP^{N-1}$ is of rank $N$, the dimension of $R^\vee$.

Moreover, the generators of the cohomology of $\CP^{N-1}$ are in degrees $0,2,4,\dots,2N-2$.   Let us shift the degrees
by $-(N-1)$ so that they are symmetrically spaced around zero.  Then we  can write a diagonal matrix whose eigenvalues
are the appropriate degrees:
\begin{equation}\label{zolme} h = \begin{pmatrix} N-1 & & & \dots & \cr
                                                                                                  & N-3 & & \dots & \cr
                                                                                                & & & \ddots & \cr
                                                                                                  & & & \dots& -(N-1)\end{pmatrix}.\end{equation}
One may recognize this matrix; it is an element of the Lie algebra of $G^\vee_\C=SL(N,\C)$ that, in the language of
Kostant,  generates the maximal torus of a ``principal $SL_2$ subgroup'' of $G^\vee_\C$.

This is the general state of affairs. In the correspondence between a representation $R^\vee$ and the cohomology
of the corresponding moduli space $\bN(\rho)$, the grading of the cohomology by degree corresponds to the action on
$R^\vee$ of a generator of the maximal torus of a principal $SL_2$.

\subsection{Characteristic Classes In Gauge Theory}\label{chg}
The nilpotent ``raising operator''   of the principal $SL_2$ also plays a role.  To understand this, first recall that Atiyah and Bott
\cite{AB} used gauge theory to define certain universal cohomology classes over any family of $G_\C$-bundles                                                                                                        over a Riemann surface $C$. The definition applies immediately
to $\bN(\rho)$, which parametrizes a family of holomorphic $G_\C$-bundles
over $\Bbb{CP}^1$ (Hecke modifications of a trivial bundle).

\def\Tr{\mathrm{Tr}}
If $G$ is of rank $r$, then the ring of invariant polynomials on the
Lie algebra $\mathfrak g$ of $G$ is itself a polynomial ring with
$r$ generators, say $P_1,\dots,P_r$, which we can take to be
homogeneous of  degrees $d_1,\dots,d_r$.  The relation of the $d_i$
to a principal $SL_2$ subgroup of $G_\C$ is as follows: the Lie
algebra $\mathfrak g$ decomposes under the principal $SL_2$ as a
direct sum
\begin{equation}\label{dirsum}\mathfrak g=\oplus_{i=1}^r\mathcal J_i\end{equation}
of irreducible modules $\mathcal J_i$ of dimensions $2d_i-1$.  (In
particular, therefore, $\sum_{i=1}^r(2d_i-1)={\mathrm {dim}}\,G$.)
For example, if $G=SU(N)$, then $r=N-1$; letting $\Tr$ denote an
invariant quadratic form on $\mathfrak g$, we can take the $P_i$ to
be the polynomials $P_i(\sigma)=\frac{1}{i+1}\Tr\,\sigma^{i+1}$ for
$\sigma\in\mathfrak g$ and $i=1,2,3,\dots,N-1$.  Thus,  $P_i$ is
homogeneous of degree $i+1$. As in this example, if $G$ is simple,
the smallest value of the degrees $d_i$ is always 2 and this value
occurs precisely once.  The corresponding polynomial $P$ is simply
an invariant quadratic form on the Lie algebra $\mathfrak g$.

\def\M{{\mathcal M}}
If $F$ is the curvature of a $G$-bundle  over any space $\M$, then
$P_i(F)$ is a $2d_i$-dimensional characteristic class, taking values
in $H^{2i}(\M)$.    (For topological purposes, it does not matter if
we consider $G$-bundles or $G_\C$-bundles.) Atiyah and Bott consider
the case that $\M$ parametrizes a family of $G$-bundles over a
Riemann surface $C$.  We let $\mathcal E\to \mathcal M\times C$  be
the corresponding universal $G$-bundle. (If necessary, we consider
the associated $G_{\mathrm {ad}}$ bundle and define the $P_i(F)$ as
rational characteristic classes.)   From the class $P_i(F)\in
H^{2i}(\M\times C)$, we can construct two families of cohomology
classes over $\M$. Fixing a point $c\in C$, and writing $\pi$ for
the projection $\M\times C\to \M$, we set $v_i$ to be the
restriction of $P_i(F)$ to $\M\times c$. We also set
$x_i=\pi_*(P_i(F))$.  Thus, $v_i\in H^{2d_i}(\M)$, and $x_i\in
H^{2d_i-2}(\M)$.  To summarize,
\begin{align}\label{zorkox}\notag v_i& = P_i(F)|_{\M\times c}\\
                                                            x_i& = \pi_*(P_i(F)).\end{align}

For our present purposes, we want $\M$ to be one of the families $\bar\N(\rho)$ of Hecke modifications of the trivial bundle
$U\to \CP^1$
at a specified point $p\in \CP^1$.  Taking $c$ to be disjoint from $p$, it is clear that the classes $v_i$ vanish for $\M=\bN(\rho)$ (since a Hecke modification at $p$ has no effect at $c$).
However, the classes $x_i$ are non-zero and interesting.

Multiplication by $x_i$ gives an endomorphism of $H^*(\bN(\rho))$ that increases the degree by $2d_i-2$.  It must map under
duality to an endomorphism $f_i$ of $R^\vee$ that increases the eigenvalue of $h$ (the generator of a Cartan subalgebra
of a principal $SL_2$) by $2d_i-2$.  Thus, we expect $[h,f_i]=(2d_i-2)f_i$.    Moreover, the $f_i$ must commute, since the $x_i$
do.

As noted above, the smallest value of the degrees $d_i$ is 2, which occurs precisely once.  So this construction gives
an essentially unique class\footnote{For $G=SU(N)$, $x$
 can
also be constructed as the first Chern class of the ``determinant line bundle'' associated to the family $\M$ of
vector bundles over $C$.  For $G=SO(N)$ or $Sp(2N)$, $x$ can similarly be constructed as the first Chern class of a
Pfaffian line bundle.} $x$ of degree 2.   It turns out that duality maps $x$ to the nilpotent raising operator of the principal $SL_2$ subgroup of $G$
that we have already encountered (the action of whose maximal torus is dual to the grading of $H^*(\bN(\rho))$ by degree).
This being so, since the $x_i$ all commute with $x$, duality must map them to elements of $\mathfrak g$ that commute with the raising operator
of the principal $SL_2$.  These are precisely
 the highest weight vectors in the $SL_2$ modules $\mathcal J_i$ of eqn. (\ref{dirsum}).

For example, for $SL(N,\C)$, the raising operator of the principal $SL_2$ is the matrix
 \begin{equation}\label{thel} f = \begin{pmatrix} 0 & 1 & 0 & \dots & 0 \cr
                                                                  0 & 0 & 1 & \dots & 0 \cr
                                                                       & & &  \ddots &  \cr
                                                                       0 & 0 & 0 & \dots & 0 \end{pmatrix},\end{equation}
                                                                       with $1$'s just above the main diagonal.
The image of the two-dimensional class $x$ under duality is
precisely $f$.   Indeed, in this example, the cohomology of
$\bN(\rho)=\CP^{N-1}$ is spanned by the classes $1,x,x^2,\dots
,x^{N-1}$, and in this basis (which we have used in writing the
degree operator as in eqn. (\ref{zolme})), $x$ coincides with the
matrix $f$. The traceless matrices that commute with $f$ (in other
words, the highest weight vectors of the $\mathcal J_i$) are the
matrices $f^k$, $k=1,\dots,N-1$. The invariant polynomial $P_i
=\frac{1}{i+1}\Tr\,\sigma^{i+1}$, with $d_i=i+1$, is associated with
a class $x_i$ of dimension $2(d_i-1)=2i$. This class maps under
duality to $f^i$.
\def\T{{\mathcal T}}

Let $\T$  be the subgroup of $G^\vee_\C$ generated by a maximal
torus in a principal $SL_2$ subgroup together with the highest
weight vectors in the decomposition (\ref{dirsum}).   A summary of
part of what we have said is that the action of $\T$ on the
representation $R^\vee$ corresponds to a natural $\T$ action on the
cohomology of $\bN(\rho)$. The rest of the $G^\vee$ action on
$R^\vee$ does not have any equally direct meaning in terms of the
cohomology of $\bN(\rho)$.

There are additional facts of a similar nature.  Our goal here is not to describe all such facts but to explain how such
facts can emerge from gauge theory.

 \subsection{Convolution And The Operator Product Expansion}\label{conv}
One obvious gap in what we have said so far is that we have treated independently each representation $R^\vee$ of the dual group $G^\vee$. For example, in section \ref{chg}, the characteristic class $x$ was defined uniformly for all $\bN(\rho)$
by a universal gauge theory construction.   But it might appear from the analysis that $x$ could map under duality to
a multiple of the Lie algebra element $f$ of equation (\ref{thel}) (or more exactly, a multiple of the linear transformation
by which $f$ acts in the representation $R^\vee$), with a different multiple for every representation.

Actually, the different irreducible representations of $G^\vee$ are linked by the classical operation of taking a tensor
product of two representations and decomposing it in a direct sum of irreducibles.  This operation is dual to a certain
natural ``convolution'' operation \cite{G,MV} on the cohomology of the moduli spaces $\bN(\rho)$. This operation also has a gauge
theory interpretation, as we recall shortly.

Suppose that $R_\alpha^\vee$ and $ R_\beta^\vee$ are two irreducible representations of $G^\vee$, and that the decomposition of their tensor product is
\begin{equation}\label{golly} R_\alpha^\vee\otimes R_\beta^\vee =\oplus_\gamma N_{\alpha\beta}^\gamma \otimes R^\vee_\gamma,\end{equation}
where $R^\vee_\gamma$ are inequivalent irreducible representations of $G^\vee$, and $N_{\alpha\beta}^\gamma$ are vector
spaces (with trivial action of $G^\vee$).
For each $\alpha$, let  $\rho_\alpha:U(1)\to T\subset G$ be the homomorphism corresponding to $R^\vee_\alpha$.
Under the duality maps $R_\alpha^\vee\leftrightarrow H^*(\bN(\rho_\alpha))$,  eqn. (\ref{golly}) must correspond to a decomposition
\begin{equation}\label{olly} H^*(\bN(\rho_\alpha))\otimes H^*(\bN(\rho_\beta)) =\oplus_\gamma N^\gamma_{\alpha\beta}
H^*(\bN(\rho_\gamma)),\end{equation}
with the same vector spaces $N^\gamma_{\alpha\beta}$ as before. Indeed  \cite{G,MV}, the appropriate decomposition can be described directly in terms of the affine Grassmannian  of $G$ without reference to duality.   This decomposition is
compatible with the action of group $\T$ described in section \ref{chg}.  In other words, $\T$ acts on each factor
on the left of (\ref{olly}), and hence on the tensor product; it likewise acts on each summand on the right of (\ref{olly})  and hence on the direct sum; and
these actions agree.

In gauge theory terms, the classical tensor product of representations (\ref{golly}) corresponds to the operator product expansion for
Wilson operators, and the corresponding decomposition (\ref{olly}) corresponds to the operator product expansion for
't Hooft operators.  This has been
 explained in  \cite{KW}, section 10.4, and that story will not be repeated here.  However,  in section \ref{fusion},
 we will  explain why the operator product expansions (or in other words the above decompositions) are compatible
 with the action of $\T$.

\def\B{\widehat B}
\def\A{\widehat A}
\def\AA{\mathcal A}
\section{Gauge Theory}\label{gaugetheory}

\subsection{The $\widehat{\cmmib A}-$ and $\widehat{\cmmib B}-$Models}\label{ab}
Let $M$ be a four-manifold.  We will be studying gauge theory on $M$
-- more specifically, the twisted version of $\NN=4$ super
Yang-Mills theory that is related to geometric Langlands duality. We
write $A$ for the gauge field, which is a connection on a bundle
$E\to M$, and $F$ for its curvature.  Another important ingredient
in the theory is a one-form $\phi$ that is valued in
$\mathrm{ad}(E)$.

As explained in \cite{KW}, the twisting introduces
an asymmetry between $G^\vee$ and $G$.  $A$ and $\phi$ combine together in quite different ways in the two cases.

In the $G^\vee$ theory, which we will loosely call the $\B$-model (because on compactification to two dimensions,
it reduces to an ordinary $B$-model), $A$ and $\phi$ combine together to a
 complexified connection $\AA=A+i\phi$.
As explained in \cite{KW},  supersymmetry in the $\B$-model requires the connection $\AA$ to be flat.
So the $\B$-model involves the study of representations of the fundamental group of $M$ in $G^\vee_\C$.
As long as the flat connection $\AA$
is irreducible, it is the only important variable in the $\B$-model.  However, we will later analyze a situation in which the
condition of irreducibility is not satisfied,
so we will encounter other variables (also described in \cite{KW}).

\def\d{{\mathrm d}}
In the $G$ theory, the pair $(A,\phi)$ instead obey a nonlinear elliptic equation
\begin{equation}\label{yerf}F-\phi\wedge\phi = \star \d_A\phi,\end{equation}
where $\star$ is the Hodge star operator and $\d_A$ is the gauge-covariant extension of the exterior derivative. This equation is analogous to the instanton equations of two-dimensional
$A$-models (as well as to other familiar equations such as Hitchin's equations in two dimensions), so we will call the model
the $\A$-model.  The equation (\ref{yerf}) may  be unfamiliar, but it has various specializations that are more familiar.  For
example, suppose that $M=W\times \R$, and that the solution is invariant under rigid motions of $\R$ including
those that reverse orientation. (To get a symmetry of (\ref{yerf}), orientation reversal must be accompanied by a sign change of $\phi$.)  Parametrizing $\R$ by a real coordinate $t$, the conditions imply that
 $A$ is pulled back from $W$, and that $\phi=\phi_0\,\d t$
where the section $\phi_0$ of $\mathrm{ad}(E)$ is also pulled back from $W$. Then (\ref{yerf})  specializes to the three-dimensional Bogomolny equations:
\begin{equation}\label{bog}F=\star \d_A\phi_0.\end{equation}
Here $\star$ is now the Hodge star operator in three dimensions. Similarly (\ref{yerf}) can be reduced to Hitchin's equations
in two dimensions.  (For this, we take $M=\Sigma\times C$, where $\Sigma$ and $C$ are two Riemann surfaces,
and assume that $A$ and $\phi$ are pulled back from $C$.)   The Bogomolny equations have been extensively studied,
for instance in \cite{AH}.

 \subsection{Wilson And 'T Hooft Operators}
Let $L\subset M$ be an embedded oriented one-manifold.  We want to make some modification along $L$ of gauge theory
on $M$.

Starting with the $\B$-model, one ``classical'' modification is to suppose that $L$ is the trajectory of a ``charged particle" in the representation $R^\vee$ of
the gauge group, which we take to be $G^\vee$.  Mathematically, we achieve this by including in the ``path integral'' of the theory a factor consisting of the trace, in the $R^\vee$ representation, of the holonomy around $L$ of the complexified connection $\AA$.  This trace might be denoted as
$\Tr_{R^\vee}\,\mathrm{Hol}(\AA,L)$; physicists usually write it as
$\Tr_{R^\vee}\,P\exp\left(-\oint_L\AA\right)$.  Since it is just a function of $\AA$, this operator preserves the topological
invariance of the $\B$-model. (The $\B$-model condition that $\AA$ is flat means that the holonomy only depends on the
homotopy class  of $L$.)  When included as a factor in a quantum path integral, the holonomy is known as a Wilson
operator.

In taking the trace of the holonomy, we have assumed that $L$ is a closed one-manifold, that is, a circle.  If $M$ is compact, this is the only relevant case.  More generally, if $M$ has boundaries or ends, one also considers the case that $L$ is an open one-manifold that connects boundaries or ends of $M$.  Then instead of a trace, one considers the matrix elements of the holonomy between
prescribed initial and final states -- that is, prescribed initial and final vectors in $R^\vee$.  This is actually the situation that
we will consider momentarily.

What is the dual in $G$ gauge theory of including the holonomy factor in $G^\vee$ gauge theory?  The dual
is the 't Hooft operator. It was essentially shown by 't Hooft nearly thirty years ago that the dual operation to including
a holonomy factor or Wilson operator  is to
modify the theory by requiring the fields to have a certain type of singularity along $L$.
This singularity gives a way to study via gauge theory the Hecke modifications of a $G$-bundle on a Riemann
surface.   The required singularity and its interpretation in terms of Hecke modifications have been described in sections 9 and 10 of \cite{KW}; a few relevant points are summarized in section \ref{bogsing}.

\subsection{Choice of $\cmmib M$}\label{choice}
In this article, our interest is in the representation $R^\vee$, not the four-manifold $M$.  So we simply want to
pick $M$ and the embedded one-manifold $L$ to be as simple as possible.  It is convenient to take $M=W\times \R$,
where $W$ is a three-manifold and $\R$ parametrizes the ``time.''    We similarly take $L=w\times \R$, where $w$ is a point
in $W$.

Henceforth we adopt  a ``Hamiltonian'' point of view in which, in
effect, we work at time zero and only talk about $W$.  So instead of
a four-manifold $M$ with an embedded one-manifold $L$ labeled by a
representation $R^\vee$, we consider a three-manifold $W$ with an
embedded point $w$ labeled by that representation.   The presence of
this special point means that, in the quantization, we must include
an ``external charge'' in the representation $R^\vee$,

Moreover, we want to make a simple choice of $W$ so as to study the representation $R^\vee$, and
its dual, keeping away from the wonders of three-manifolds.

What is the simplest three-manifold?  $S^3$ comes to mind right away, but there is a snag.   Suppose that we study
the $G^\vee$ gauge theory on $W=S^3$, with a marked point $w$ that is labeled by the representation $R^\vee$.
What will the quantum Hilbert space turn out to be?  A flat connection on $S^3$ is necessarily trivial, so there is no moduli
space of flat connections to quantize.  If the trivial flat connection on $S^3$ had no automorphisms, the quantum Hilbert space
of the $\B$-model would be simply $R^\vee$, as there is nothing else
to quantize.  However, the trivial flat connection on $S^3$ actually has a group $G_\C$ of automorphisms, and in quantization, one is supposed to impose invariance under the group
of gauge transformations.  Because of this, the quantum Hilbert space is not $R^\vee$ but the $G^\vee_\C$ invariant subspace of $R^\vee$ -- namely
0.  Thus, simply taking $W=S^3$, with a marked point labeled by the representation $R^\vee$, will not give us a way
to use the $\B$-model to study the representation $R^\vee$.

What we seem to need -- a three-manifold on which the trivial flat connection is unique and irreducible -- does not exist.
However, we can pick $W$ to be a three-manifold with boundary, provided that we endow the boundary with a supersymmetric
boundary condition.  For example, suppose that $W$ is a three-dimensional
ball $B^3$.  We may pick Dirichlet boundary conditions on the boundary of $B^3$.  In the $\B$-model, Dirichlet boundary
conditions mean that $\AA$ is trivial on the boundary $\partial B^3$, and that only gauge transformations that are trivial
on the boundary are allowed.

If we formulate the $\B$-model on $B^3$ with Dirichlet boundary
conditions and a marked point labeled by $R^\vee$, then as there are
no nontrivial flat connections and the trivial one has no gauge
symmetries, the physical Hilbert space is a copy of $R^\vee$.  So
this does give a way to study the representation $R^\vee$ as a space
of physical states in the $\B$-model.  The only trouble is that the
dual of Dirichlet  boundary conditions is rather complicated
\cite{GW1,GW2}, and the resulting $\A$-model picture is not very
transparent.

There is another choice that turns out to be more useful because it gives something that is tractable in both the
$\A$-model and the $\B$-model.  This is to take $W=S^2\times I$, where $I\subset \R$ is a closed interval.  Of course, $W$ has two ends, since $I$ has two boundary points.  Suppose that we pick Dirichlet boundary conditions at one end of $S^2\times I$
and Neumann boundary conditions at the other.  (Neumann boundary conditions in gauge theory mean that the gauge
field and the gauge transformations are arbitrary on the boundary; instead there is a condition on the normal derivative
of the gauge field, though we will not have to consider it explicitly because  it is a consequence of the equations that we will be solving anyway.)  With these boundary conditions, the trivial flat connection on $W$ is unique and irreducible.

By contrast, if we were to place Dirichlet boundary conditions at both ends of $S^2\times I$, there would be non-trivial flat connections classified by the holonomy along a path from one end to the other; with Dirichlet boundary conditions at both
ends, this holonomy is gauge-invariant.  And if we were to place Neumann boundary conditions at both ends, every flat
connection would be gauge-equivalent to the trivial one, but (since there would be no restriction on the boundary
values of a gauge transformation) the trivial flat connection would have a group $G_\C$ of
automorphisms, coming from constant gauge transformations.

The case that works well is therefore the case of  mixed boundary conditions -- Dirichlet at one end and Neumann at the other.
So we could study the representation $R^\vee$ in the $\B$-model by working on $W=S^2\times I$ with mixed
boundary conditions
 This may even be an interesting thing to do.

\begin{figure}
  \begin{center}
    \includegraphics[width=3in]{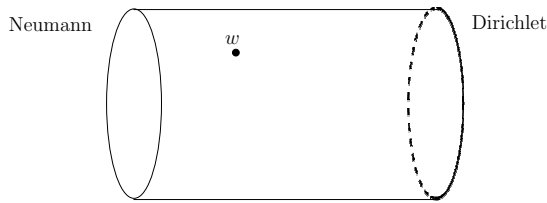}
  \end{center}
\caption{{{ Depicted here is $W=S^2\times I$ in $G$ gauge theory
with a marked point $w$ at which an 't Hooft operator is inserted.
Dirichlet boundary conditions are imposed at the right boundary and
Neumann boundary conditions at the left. }}}
  \label{octo}\end{figure}

Instead here, we will do something that turns out to be simpler.  We
will study the $\A$-model, not the $\B$-model, on $W=S^2\times I$,
with mixed Dirichlet and Neumann boundary conditions, and one marked
point labeled by an 't Hooft singularity (see fig. \ref{octo}).
Since they make the trivial solution of the Bogomolny equations
isolated and irreducible, mixed boundary conditions simplify the
$\A$-model just as they simplify the $\B$-model.

In fact, the $\A$-model on $S^2\times I$ with mixed boundary conditions was studied in \cite{KW}, section 10.4, in order
to investigate the operator product expansion for 't Hooft operators.
  At the time, it was not possible to compare
to a $\B$-model description, since the duals of Dirichlet and Neumann boundary conditions in supersymmetric non-abelian
gauge theory were not sufficiently clear.  Here, we will  complete the analysis using more recent results \cite{GW1,GW2} on duality of boundary conditions.  This will enable us to
understand via gauge theory the results that were surveyed in the introduction.

\subsection{Bogomolny Equations With A Singularity}\label{bogsing}

In the $\A$-model on $W=S^2\times I$, we must solve the Bogomolny equations (\ref{bog}), with singularities
at the positions of 't Hooft operators.  For the moment, suppose that there is a single such singularity, located at
$w=c\times r$, where $c$ and $r$ are points in $S^2$ and $I$, respectively.

If $E$ is any $G$-bundle with connection over $C\times I$, where $C$ is a Riemann surface (in our case, $C=S^2$), we can
restrict $E$ to $C\times \{y\}$, for $y\in I$, to get a $G$-bundle with connection $E_y\to C$.  Since any connection on  a bundle
over a Riemann surface defines an integrable $\bar\partial$ operator, the bundles $E_y$ are in a natural way holomorphic
$G_\C$-bundles.

One of the many special properties of the Bogomolny equations is
that if the pair $(A,\phi_0)$ obeys these equations, then as a
holomorphic bundle, $E_y$ is independent of $y$, up to a natural
isomorphism.  This is proved by a very short computation.  Writing
$z$ for a local holomorphic coordinate on $C$, a linear combination
of the Bogomolny equations gives $F_{y\bar z}=-iD_{\bar z}\phi_0$,
or $[\partial_y+A_y+i\phi_0,\bar\partial_A]=0$, where
$\bar\partial_A$ is the $\bar\partial$ operator on $E_y$ determined
by the connection $A$.  Thus, $\bar\partial_A$ is independent of
$y$, up to a complex gauge transformation, and integrating the
modified connection $A_y+i\phi_0$ in the $y$ direction gives a
natural isomorphism between the $E_y$ of different $y$.

In the presence of an 't Hooft operator at $w=c\times r$, the Bogomolny equations fail (because there is a singularity) at the
point $w$, and as a result, the holomorphic type of $E_y$ may jump when we cross $y=r$.  However, if we delete from $C$ the point $c$, then we do not see the singularity and no jumping occurs.  In other words, if we write $E'_y$ for the restriction of $E_y$
to $C\backslash c$, then $E'_y$ is independent of $y$, as a holomorphic bundle over $C\backslash c$.  (Moreover, there is
a natural isomorphism between the $E'_y$ of different $y$, by parallel transport with the connection $A_y+i\phi_0$.)
Thus, the jump in $E_y$ in crossing $y=r$ is a Hecke modification at the point $c\in C$.

Suppose that $R^\vee$ is an irreducible representation of the dual group $G^\vee$.
Using ideas described in section \ref{dualg}, let $\rho:U(1)\to G$ be the  homomorphism corresponding to $R^\vee$,
and $E_\rho\to \CP^1$ the corresponding $G_\C$-bundle.    Then the 't Hooft operator dual to $R^\vee$ in $G$ gauge theory is defined so that the Hecke modification found in the last paragraph is of type $\rho$.   This is accomplished by specifying a suitable singularity type in the solution of the Bogomolny equations.
Roughly, one arranges so that the solution $(A,\phi_0)$ of the Bogomolny equations has the property that,
when restricted to a small two-sphere $S$ that encloses the point $w$, the connection $A$ determines a holomorphic
$G_\C$-bundle over $S$ that is equivalent holomorphically to $E_\rho$.  For details, see \cite{KW}.

\subsection{The Space of Physical States}\label{spacest}

Now let us determine the space of physical states of the $\A$-model on $W=S^ 2\times I$, with mixed boundary conditions.
On general grounds, this is the cohomology of the moduli space of solutions of the Bogomolny equations, with the chosen
boundary conditions.

Dirichlet boundary conditions  at one end of $W$ means that $E_y$ is
trivial at that end.   Neumann boundary conditions means that, at
the other end, any $E_y$ that is produced by solving the Bogomolny
equations is allowed.   In the presence of a single 't Hooft
operator dual to $R^\vee$, any Hecke modification of type $\rho$ can
occur.  So the moduli space of solutions of the Bogomolny equations
is our friend, the moduli space $\N(\rho)$ of Hecke modifications of
type $\rho$. This moduli space has a natural compactification by
allowing monopole bubbling \cite{KW,CD}, the shrinking to a point of
a lump of energy in a solution of the Bogomolny
equations.\footnote{Monopole bubbling is somewhat analogous to
instanton bubbling in four dimensions, which involves the shrinking
of an instanton.  An important difference is that instanton bubbling
can occur anywhere, while monopole bubbling can only occur at the
position of an 't Hooft operator.  Monopole bubbling involves a
reduction of the weight $\rho$ associated to the 't Hooft
singularity.} This compactification is the compactified space
$\bN(\rho)$ of Hecke modifications.

The space $\mathcal H$ of physical states of the $\A$-model is therefore the cohomology $H^*(\bN(\rho))$.  Together
with the fact that we will find $\mathcal H=R^\vee$ in the $\B$-model, this is the basic
reason that electric-magnetic duality establishes a map between the cohomology $H^*(\bN(\rho))$
and the representation $R^\vee$ of $G^\vee$.

\subsection{Nahm's Equations}\label{backtogvee}

To learn more about  $\mathcal H$, we need to analyze its dual description in $G^\vee$ gauge theory.
Some things are simpler than what we have met so far, and some things are less simple.

First of all, there are no Bogomolny equations to worry about.  The supersymmetric equations of the $\B$-model
are quite different.  As formulated in \cite{KW}, these equations involve a connection
$A$ on a $G^\vee$ bundle $E^\vee\to M$, a one-form $\phi$ valued in  the adjoint bundle $\mathrm{ad}(E^\vee)$,
and a zero-form $\sigma^\vee$ taking values in the complexification $\mathrm{ad}(E^\vee)\otimes\C$.  (We write $\sigma^\vee$
for this field -- a slight departure from the notation in \cite{KW} -- as we will later introduce an analogous field $\sigma$
in the $\A$-model.) It is convenient
to combine $A$ and $\phi$ to a complex connection
$\AA=A+i\phi$ on the $G_\C^\vee$-bundle $E_\C^\vee\to M$ obtained by complexifying $E^\vee$.  Moreover, we
write $\mathcal F$ for the curvature of $\AA$, and $\d_\AA$, $\d_A$ for the exterior derivatives with respect to $\AA$ and $A$,
respectively.
The supersymmetric conditions read
\begin{align}\label{donkey}\notag   \mathcal F_\AA & = 0 \\
                                                       \d_\AA\sigma^\vee & = 0 \\ \notag
                                                      \d_A^*\phi+i[\sigma^\vee,\bar\sigma^\vee] &= 0 . \end{align}
Here
$\d_A^*=\star \d_A\star$ is the adjoint of $\d_A$.
The first condition says that $E_\C^\vee$ is flat, and the second condition says that $\sigma^\vee$ generates an automorphism
of this flat bundle.  If therefore $E_\C^\vee$ is irreducible, then $\sigma^\vee$ must vanish.  This is the case most often considered
in the geometric Langlands correspondence, but we will be in a rather different situation because, for $W=S^2\times I$ and
with the boundary conditions we have introduced, there are no non-trivial flat connections.   While the first
two equations are invariant under $G^\vee_\C$-valued gauge transformations, the third one is only invariant under $G^\vee$-valued gauge transformations.  For a certain natural symplectic structure on the data $(A,\phi,\sigma^\vee)$, the expression
$\d_A^\star\phi+i[\sigma^\vee,\bar\sigma^\vee]$ is the moment map for the action of $G^\vee$ gauge transformations on this data.  As this interpretation suggests,
the third equation is a stability condition; the moduli space of solutions of the three equations, modulo $G^\vee$-valued  gauge
transformations, is the moduli space of stable pairs $(\AA,\sigma^\vee)$ obeying the first two equations, modulo $G^\vee_\C$-valued
gauge transformations.  A pair is considered strictly stable if it cannot be put in a triangular form
\begin{equation}\label{tri}\begin{pmatrix}\alpha & \beta \\ 0 & \gamma\end{pmatrix},\end{equation}
and semistable if it can be put in such a form. (There are no strictly unstable pairs.) Two semistable pairs are considered equivalent if the diagonal blocks
$\alpha$ and $\gamma$ coincide.    For the case $\sigma^\vee=0$, this interpretation of the
third equation was obtained in \cite{C}.

Rather surprisingly, the system of equations (\ref{donkey}) can be
truncated to give a system of equations in mathematical physics that
are familiar but are not usually studied in relation to complex flat
connections.  These are Nahm's equations. They were originally
obtained \cite{N} as the result of applying an ADHM-like transform
to the Bogomolny equations on $\R^3$; subsequently, they have turned
out to have a wide range of mathematical applications, for instance
see \cite{Kr,AtBi}.

To reduce the equations (\ref{donkey}) to Nahm's equations, suppose that $A=0$ and that $\phi=\phi_y\,\d y$,
where $y$ is one of the coordinates on $M$.  In our application, we have $M=W\times \R$, $W=S^2\times I$, and we take
$y$ to be a coordinate on $I$, so that $y=0$ is one end of $I$.  Furthermore, write $\sigma^\vee=(X_1+iX_2)/\sqrt 2$, where $X_1,X_2$
take values in the {\it real} adjoint bundle $\mathrm{ad}(E^\vee)$, and set
\begin{equation}\label{egg}\phi_y=X_3.\end{equation}  Then the equations (\ref{donkey})
reduce unexpectedly to Nahm's equations $\d X_1/\d y+[X_2,X_3]=0$, and cyclic permutations of indices $1,2,3$.  Alternatively, combining
$X_1,X_2,X_3$ to a section $\vec X$ of $\mathrm {ad}(E^\vee)\otimes \R^3$, the equations can be written
\begin{equation}\label{logic}\frac{\d\vec X}{\d y}+\vec X\times \vec X = 0.\end{equation}
Here $(\vec X\times \vec X)_1=[X_2,X_3]$, etc.

Nahm's equations (\ref{logic}) have an obvious $SO(3)$ symmetry acting on $\vec X$.   In the way we have derived these
equations from (\ref{donkey}),  this symmetry is rather mysterious.  Its origin is more obvious in the underlying four-dimensional
gauge theory, as we explain in section \ref{details}.

\subsection{The Dual Boundary Conditions}\label{subtlety}

\def\Spin{\mathrm{Spin}}
Nahm's equations admit certain singular solutions that are important
in many of their applications \cite{N,Kr,AtBi}. Let
$\vartheta:\mathfrak{su}(2)\to \mathfrak g^\vee$ be any homomorphism
from the $SU(2)$ Lie algebra to that of $G^\vee$. It is given by
elements $\vec t=(t_1,t_2,t_3)\in \mathfrak g^\vee$ that obey the
$\mathfrak{su}(2)$ commutation relations $[t_1,t_2]=t_3$, and cyclic
permutations.  Then Nahm's equations on the half-line $y>0$ are
obeyed by
\begin{equation}\label{zonko}\vec X=\frac{\vec t}{y}.\end{equation}

Consider $G^\vee$ gauge theory on a half-space $y\geq 0$. Dirichlet
boundary conditions on $G^\vee$ gauge fields can be extended to the
full $\NN=4$ super Yang-Mills theory in a supersymmetric (half-BPS)
fashion.  When this is done in the most obvious way, the fields
$\vec X$ actually obey free (or Neumann) boundary conditions and
thus are unconstrained, but nonsingular, at the boundary. With the
aid of the singular solutions (\ref{zonko}) of Nahm's equations, one
can describe boundary conditions  \cite{GW1} in $G^\vee$ gauge
theory that generalize the most obvious Dirichlet boundary
conditions in that they preserve the same supersymmetry. To do this,
instead of saying that $\vec X$ is regular at $y=0$, we say that it
should have precisely the singular behavior of (\ref{zonko}) near
$y=0$. This condition can be uniquely extended to the full $\NN=4$
theory in a supersymmetric fashion. This use of a classical
singularity to define a boundary condition in quantum theory is
somewhat analogous to the definition of the 't Hooft operator via a
classical singularity (in that case, a singularity along a
codimension three submanifold of spacetime).

The most important case for us
 will be what we call a regular Nahm pole.  This is the case that $\vartheta:\frak{su}(2)\to
\frak g^ \vee$ is a principal embedding.  (Usually the principal
embedding is defined as a homomorphism $\frak{sl}(2,\C)\to \frak
g^\vee_\C$; the complexification of $\vartheta$ is such a
homomorphism.) For $G^\vee_\C=SL(N,\C)$, a principal $\frak{sl}(2)$
embedding (or at least the images of two of the three
$\mathfrak{sl}(2)$ generators) was described explicitly in eqns.
(\ref{zolme}) and (\ref{thel}).  As in this example, a principal
$\mathfrak{sl}(2)$ embedding $\vartheta$ is always irreducible in
the sense that the subalgebra of $\mathfrak g_\C^\vee$ that commutes
with the image of $\vartheta$ is zero.  Conversely, an irreducible
$\mathfrak{sl}(2)$ embedding is always conjugate to a principal one.

\subsubsection{The Dual Picture}
Finally we have the tools to discuss the dual of the $\A$-model
picture that was analyzed in section \ref{spacest}. In our study of
$G$ gauge theory, we imposed mixed Dirichlet-Neumann boundary
conditions -- say Neumann at $y=0$, and Dirichlet at $y=L$. To
compare to a description in $G^\vee$ gauge theory, we need to know
what happens to Neumann and Dirichlet boundary conditions under
duality.

For $G=G^\vee=U(1)$, electric-magnetic duality simply exchanges
Dirichlet and Neumann boundary conditions.  One of the main results
of \cite{GW1,GW2} is that this is not true for nonabelian gauge
groups.  Rather, electric-magnetic duality maps Neumann boundary
conditions to Dirichlet boundary conditions modified by a regular
Nahm pole.   And it maps Dirichlet boundary conditions to something
that is very interesting (and related to the ``universal kernel'' of
geometric Langlands, as explained in section \ref{genker}) but more
difficult to describe.

For our purposes, all we really need to know about the dual of
Dirichlet boundary conditions is that in the $\B$-model on
$W=S^2\times I$ (times $\R$), with boundary conditions at $y=0$
given by the regular Nahm pole, and the appropriate boundary
conditions at $y=L$, the solution of Nahm's equations is unique.  In
fact, the relevant solution is precisely $\vec X=\vec t/y$.  (The
boundary conditions at $y=0$ require that the solution should take
this form, modulo regular terms; the regular terms are fixed by the
boundary condition at $y=L$.)  How this comes about is described in
section \ref{realdetails}.

\begin{figure}
  \begin{center}
    \includegraphics[width=4in]{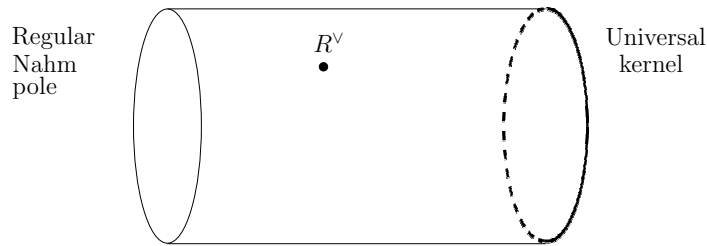}
  \end{center}
\caption{{  $W=S^2\times I$ in $G^\vee$ gauge theory with a marked
point at which is included an external charge in the representation
$R^\vee$. The boundary conditions are dual to those of fig.
\ref{octo}. At the left are Dirichlet boundary conditions modified
with a regular Nahm pole, while at the right are more complicated
boundary conditions associated with the universal kernel of
geometric Langlands.}}
  \label{bocto}\end{figure}

The dual picture, therefore, is as described in fig. \ref{bocto}.

\subsection{The Space Of Physical States In The $\widehat{\cmmib
B}$-Model}\label{space} Now we can describe the space of physical
states in the $\B$-model on  $S^2\times I$, with these boundary
conditions, and with a marked point $w=c\times r$ labeled by the
representation $R^\vee$.

The analysis is easy because with the boundary conditions, Nahm's equations have a unique and irreducible solution, with no gauge automorphisms and no moduli
that must be quantized.    Moreover, no moduli appear when Nahm's equations are embedded in the more complete system
(\ref{donkey}).   This follows from the irreducibility of the solution of Nahm's equations with a regular pole.

In the absence of marked points, the physical Hilbert space
$\mathcal H$ would be a  copy of $\C$, from quantizing a space of
solutions of Nahm's equations that consists of only one point.
However, we must take into account the marked point.  In general, in
the presence of the marked point, $\mathcal H$ would be computed as
the $\bar\partial$ cohomology of a certain holomorphic vector
bundle\footnote{Let $\mathcal E\to \M\times W$ be the universal
bundle. We construct the desired bundle $\mathcal E_{R^\vee}\to\M$
by restricting $\mathcal E$ to $\M\times w$ and taking the
associated bundle in the representation $R^\vee$. (In this
construction, in general $\mathcal E$ must be understood as a
twisted bundle, twisted by a certain gerbe.)} with fiber $R^\vee$
over the moduli space $\mathcal M$ of solutions of the equations
(\ref{donkey}). (If a generic point in $\mathcal M$ has an
automorphism group $H$, then one takes the $H$-invariant part of the
$\bar\partial$ cohomology.) In the present case, as $\mathcal M$ is
a single point (with no automorphisms), the physical Hilbert space
is simply $\mathcal H= R^\vee$.

Therefore, electric-magnetic duality gives a natural map from $H^*(\bN(\rho))$, which is the space $\mathcal H$ of physical states computed in the $\A$-model, to $R^\vee$.  Now we can try to identify in the $\B$-model the grading of $\mathcal H$ that
in the $\A$-model corresponds to the grading of the cohomology $H^*(\bN(\rho))$ by degree.

In the underlying $\NN=4$ super Yang-Mills theory, there is a
$\Spin(6)$ group of global symmetries (these symmetries act
non-trivially on the supersymmetries and are hence usually called
$R$-symmetries). The twisting that leads eventually to geometric
Langlands breaks this $\Spin(6)$ symmetry down to $\Spin(2)$. (This
remark and related remarks in the next paragraph are explained more
fully in  section \ref{details}.) In the context of topological
field theory, this $\Spin(2)$ symmetry is usually called ``ghost
number.''  The action of this $\Spin(2)$ symmetry on the $\A$-model
gives the grading of $H^*(\bN(\rho))$ by degree.

So we must consider the action of the $\Spin(2)$ or ghost number
symmetry in the $\B$-model.  In the $\B$-model, $\Spin(2)$ acts by
rotation of $\sigma^\vee$, that is, by rotation of the $X_1-X_2$
plane.  To be more precise, $\sigma^\vee$ has ghost number 2;
equivalently, the $\Spin(2)$ generator acts on the $X_1-X_2$ plane as
$2(X_1\partial/\partial X_2-X_2\partial/\partial X_1)$.

In quantizing the $\B$-model on $S^2\times I$ with the boundary
conditions that we have chosen, $X_1$ and $X_2$ are not zero -- in
fact, they appear in the solution of Nahm's equations with the
regular pole, $\vec X=\vec t/y$. So this solution is not invariant
under a rotation of the $X_1-X_2$ plane, understood naively.   Why
therefore in the $\B$-model is there a $\Spin(2)$ grading of the
physical Hilbert space $\mathcal H$?

The answer to this question is that we must accompany a  $\Spin(2)$
rotation of the $X_1-X_2$ plane with a gauge transformation. The
regular Nahm pole $\vec X=\vec t/y$ is invariant under the
combination of a rotation of the $X_1-X_2$ plane and a gauge
transformation generated by $t_3$.   The rotation of the $X_1-X_2$
plane does not act on the representation $R^\vee$, but the gauge
transformation does. So on the $\B$-model side, the grading of
$\mathcal H$ comes from the action of $t_3$. But since the boundary
condition involves a regular Nahm pole, $t_3$ generates the maximal
torus of a principal $SL_2$ subgroup of $G^\vee$.

So electric-magnetic duality maps the grading of $H^*(\bN(\rho))$ by
degree to the action on $R^\vee$ of the maximal torus of a principal
$SL_2$ subgroup. This fact was described in section \ref{principal}.
Now we understand it via gauge theory.

\subsection{Universal Characteristic Classes In The $\widehat{\cmmib A}$-Model}\label{universal}

It remains to understand via gauge theory an additional fact
described in section \ref{chg}: under duality, certain natural
cohomology classes of $\bN(\rho)$ map to elements of $\mathfrak
g^\vee$ acting on $R^\vee$. There are three steps in understanding
this: (i) interpret these cohomology classes as local quantum field
operators in the $\A$-model; (ii) determine their image under
electric-magnetic duality; (iii) compute the action of the dual
operators in the $\B$-model. We consider step (i) here and steps
(ii) and (iii) in section \ref{othersteps}.

How to carry out step (i) is known from experience with Donaldson
theory. In defining  polynomial invariants of four-manifolds
\cite{D}, Donaldson adapted to four dimensions the universal gauge
theory cohomology classes that in two dimensions were described in
\cite{AB} (and reviewed in section \ref{chg}). Donaldson's
construction was interpreted in quantum field theory in \cite{EWT}.
One of the main steps in doing so was to interpret the universal
characteristic classes in terms of quantum field theory operators.
The resulting formulas were understood geometrically by Atiyah and
Jeffrey \cite{AJ}. Formally, the construction of Donaldson theory by
twisting of $\NN=2$ super Yang-Mills theory is just analogous to the
construction of the $\A$-model relevant to geometric Langlands by
twisting of $\NN=4$ super Yang-Mills theory. (The instanton equation
plays the same formal role in Donaldson theory that the equation
$F-\phi\wedge \phi=\star \d_A\phi$ plays in the $\A$-model related
to geometric Langlands.) As a result, we can carry out step (i) by
simply borrowing the construction of \cite{EWT}.

As in section \ref{chg}, the starting point is an invariant
polynomial $P_i$ on the Lie algebra $\frak g$ of $G$. Using this
polynomial, one constructs corresponding supersymmetric operators in
the $\A$-model (or in Donaldson theory).  The construction uses the
existence of a field $\sigma$ of degree or ghost number 2, taking
values in the adjoint bundle $\mathrm{ad}(E)_\C$ associated to a
$G$-bundle $E$.  (There is also an analogous field $\sigma^\vee$ in
the $\B$-model; it has already appeared in the equations
(\ref{donkey}), and will reappear in section \ref{othersteps}.)
$\sigma$ is invariant under the topological supersymmetry of the
$\A$-model, so it can be used to define operators that preserve the
topological invariance of that model.

\def\P{\mathcal P}
 The most obvious way to do this is simply to define
$\P_i(z)=P_i(\sigma(z))$. This commutes with the topological
supersymmetry of the $\A$-model, since it is a function only of
$\sigma$, which has this property. Here $z$ is a point in a
four-manifold $M$, and we have made the $z$-dependence explicit to
emphasize that $\P_i$ is supposed to be a local operator in quantum
field theory. We usually will not write explicitly the
$z$-dependence.

Suppose that $P_i(\sigma)$ is homogeneous of degree $d_i$.  Then, as
$\sigma$ has degree 2, $\P_i$ is an operator of degree $2d_i$.  It
corresponds to the cohomology class $v_i$ of degree $2d_i$ that was
defined from a more topological point of view in eqn. (\ref{zorkox}).
(The link between the two points of view depends on the fact that
$\sigma$ can be interpreted as part of the Cartan model of the
equivariant cohomology of the gauge group acting on the space of
connections and other data; see \cite{AJ} for related ideas.) This
has an important generalization, which physicists call the descent
procedure; it is possible to derive from the invariant polynomial
$P_i$ a family of $r$-form valued supersymmetric operators of degree
$2d_i-r$, for $r=1,\dots,4$. (The definition stops at $r=4$ since we
are in four dimensions.) Let us write
$\widehat\P_i=\P_i^{(0)}+\P_i^{(1)}+\dots+\P_i^{(4)}$, where
$\P_i^{(0)}=\P_i=P_i(\sigma)$, and $\P_i^{(r)}$ will be a local operator
with values in $r$-forms on $M$.  We define the $\P_i^{(r)}$ for
$r>0$ by requiring that
\begin{equation}\label{werq}(\d+[Q,~\cdot~])\widehat\P_i=0,\end{equation}
 where $\d$ is
the ordinary exterior derivative on $M$ and $Q$ is the generator of
the topological supersymmetry.

$\widehat \P_i$ is uniquely determined by the condition (\ref{werq})
plus the choice of $\P_i^{(0)}$ and the fact that $\widehat\P_i$ is
supposed to be a locally defined quantum field operator (in other
words, a universally defined local expression in the fields of the
underlying super Yang-Mills theory). For example, both in Donaldson
theory and for our purposes, the most important component is the
two-form component $\P_i^{(2)}$. It turns out to be
\begin{equation}\label{turnsout}\P_i^{(2)}=\biggl\langle\frac{\partial
P_i}{\partial\sigma},F\biggr\rangle+\biggl\langle\frac{\partial^2
P_i}{\partial\sigma^2},\psi\wedge\psi\biggr\rangle.\end{equation}
The notation here means the following.  As $P_i$ is an invariant
polynomial on the Lie algebra $\frak g$, we can regard $\partial
P_i/\partial\sigma$ as an element of the dual space $\frak g^*$.
Hence it can be paired with the $\frak g$-valued two-form $F$ (the
curvature of the gauge connection $A$) to make a gauge-invariant
two-form valued field that appears as the first term on the right of
eqn. (\ref{turnsout}).  Similarly, we can consider
$\partial^2P_i/\partial\sigma^2$ as an element of $\frak g^*\otimes
\frak g^*$.  On the other hand, $\psi$ is a $\frak g$-valued
fermionic one-form (of degree or ghost number 1) that is part of the
twisted super Yang-Mills theory under consideration here (either
twisted $\NN=2$ relevant to Donaldson theory, or twisted $\NN=4$
relevant to geometric Langlands).  So $\psi\wedge\psi$ is a two-form
valued in $\frak g\otimes\frak g$; it can be paired  with
$\partial^2P_i/\partial\sigma^2$ to give the second term on the
right hand side of (\ref{turnsout}).

From (\ref{werq}), we have $[Q,\P_i^{(2)}]=-\d \P_i^{(1)}$; thus
$[Q,\P_i^{(2)}]$ is an exact form.  So the integral of $\P_i^{(2)}$
over a two-cycle $\mathcal S\subset M$, that is
\begin{equation}\label{gutr}x_i(\mathcal S)=\int_{\mathcal S}\P_i^{(2)},\end{equation}
commutes with the generator $Q$ of the topological supersymmetry.
 Thus $x_i(\mathcal S)$ is an observable of
the $\A$-model. Since $\d\,\P_i^{(2)}=-\{Q,\P_i^{(3)}\}$, this
observable only depends on the homology class of $\mathcal S$.
Concretely, $x_i(\mathcal S)$ will correspond to a cohomology class
on the relevant moduli spaces.

In our problem with $M=W\times \R$, $W=S^2\times I$, and an 't Hooft
operator supported on $w\times \R$ with   $w\in W$, what choice do
we wish to make for $\mathcal S$?  Part of the answer is that we
will take $\mathcal S$ to be supported at a particular time.  In
other words, we take it to be the product of a point $t_0\in \R$ and
a two-cycle in $W$ that we will call $S$.

\remark \label{goodsy} The fact that $\mathcal S$ is localized in
time means that the corresponding quantum field theory expression
$x_i(\mathcal S)$ is an operator that acts on the quantum state at a particular time.
(In topological field theory, the precise time does not matter, but in general, as operators
may not commute, their ordering does.)
By
contrast, the 't Hooft operator in this problem is present for all
time, as its support is $w\times \R$. Being present
for all time, it is part of the definition of the quantum state,
rather than being an operator that acts on this state.
\bigskip

\begin{figure}
  \begin{center}
    \includegraphics[width=2in]{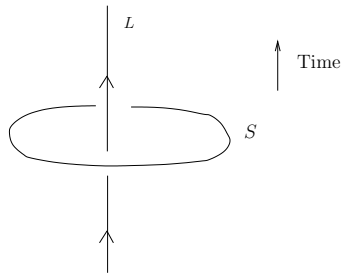}
  \end{center}
\caption{Drawn here is an 't Hooft or Wilson line operator that runs
in the time direction (shown vertically) at a fixed position in $W$.
A small two-surface $S$ (sketched here as a circle) is supported at
a fixed time and is linked with $L$.}
  \label{wocto}\end{figure}

What will we choose for $S\subset W$?  (In what follows, we will not
distinguish in the notation between $S\subset W$ and $\mathcal
S=S\times t_0\in W\times \R$.)
 One obvious choice is to let $S$ be the left or right boundary of
$W$.  If $S$ is the right boundary, where we have imposed Dirichlet
boundary conditions, so that the $G$-bundle is trivialized, then
$x_i(S)$ vanishes. (From a quantum field theory point of view, the
supersymmetric extension of Dirichlet boundary conditions actually
says that $A,\psi$, and $\sigma$ all vanish, so certainly $\widehat
\P_i$ does.) On the other hand, if $S$ is the left boundary, with
Neumann boundary conditions, there is no reason for $x_i(S)$ to
vanish.  The difference between the left and right boundaries of
$S^2\times I$ is homologous to a small two-sphere that ``links'' the
point $w=c\times r\in W$ at which an 't Hooft operator is present.
This is the most illuminating choice of $S$ (fig. \ref{wocto}).  At
any rate, whether we make this choice of $S$ or take $S$ to be
simply the left boundary of $W$, $x_i(S)$ coincides with the class
$x_i\in H^{2d_i-2}(\bN(\rho))$ that was defined in eqn.
(\ref{zorkox}).  In view of Remark \ref{goodsy}, we should think of
$x_i(S)$ as not just an element of $H^{2d_i-2}(\bN(\rho))$ but an
operator acting on this space (by cup product, as follows from
general properties of the $\A$-model).

In section \ref{chg}, we also used the invariant polynomial $P_i$ to
define gauge theory characteristic classes $v_i$ of degree $2d_i$.
As we have already mentioned, in the quantum field theory language,
these classes simply correspond to the quantum field operator
$\P_i(z)$, evaluated at an arbitrary point $z\in M$.  In section
\ref{chg}, we noted that the $v_i$ vanish as elements of
$H^{2d_i}(\bN(\rho))$ (though of course they are nonzero in other
gauge theory moduli spaces).  We can prove this in the quantum field
theory approach by taking $z$ to approach the Dirichlet boundary of
$W$; on this boundary, $\sigma=0$ so $\P_i$ vanishes.

\subsection{A Group Theory Interlude}\label{groupth}

Before describing the dual picture, we need a small group theory
interlude.

Let $T$ and $T^\vee$ be the maximal tori of $G$ and $G^\vee$ and let
$\frak t$ and $\frak t^\vee$ be their Lie algebras.  Because $\frak
t$ and $\frak t^\vee$ are dual vector spaces, and $G$ and $G^\vee$
have the same Weyl group, Weyl-invariant and nondegenerate quadratic
forms on $\frak t$ correspond in a natural way to  Weyl-invariant
and nondegenerate quadratic forms on $\frak t^\vee$. Indeed,
thinking of an invariant quadratic form $\gamma$ on $\frak t$ as a
Weyl-invariant map from $\frak t$ to $\frak t^\vee$, its inverse
$\gamma^{-1}$ is a Weyl-invariant map in the opposite direction or
equivalently a quadratic form on $\frak t^\vee$.
 If $\gamma$
and $\gamma^\vee$ are invariant quadratic forms on the Lie algebras
$\frak g$ and $\frak g^\vee$ whose restrictions to $\mathfrak t$ and
$\mathfrak t^\vee$ are inverse matrices, then we formally write
$\gamma^\vee=\gamma^{-1}$ even without restricting to $\mathfrak t$
and $\mathfrak t^\vee$.  (As we state more fully later,  the most
natural relation between quadratic forms on the two sides that comes
from duality really contains an extra factor of $n_{\mathfrak g}$,
the ratio of length squared of long and short roots.)

$G$-invariant polynomials on the Lie algebra $\frak g$ are in
natural correspondence with Weyl-invariant polynomials on $\frak t$.
Similarly $G^\vee$-invariant polynomials on $\frak g^\vee$
correspond naturally to Weyl-invariant polynomials on $\frak
t^\vee$.

Combining the above statements, once an invariant quadratic form $\gamma^\vee$ or $\gamma$ is
picked on $\frak g^\vee$ or equivalently on $\frak g$, we get a
natural map from homogeneous invariant polynomials on $\frak g$ to
homogeneous invariant polynomials on $\frak g^\vee$ of the same
degree.  Given an invariant polynomial on $\frak g$, we  restrict to a Weyl-invariant polynomial on $\frak t$,
multiply by a suitable power of $\gamma^\vee$ so it can be interpreted as a Weyl-invariant polynomial
on $\frak t^\vee$, and then associate it to a $G^\vee$-invariant polynomial on $\frak g^\vee$.
To restate this, let $({\mathrm {Sym}}^{d_i}(\mathfrak g))^G$ and
$({\mathrm{Sym}}^{d_i}({\mathfrak g}^\vee))^{G^\vee}$ be the spaces
of homogeneous and invariant polynomials of the indicated degrees,
and let $\Theta$ and $\Theta^\vee$ be the spaces of invariant
quadratic forms on $\mathfrak g$ and $\mathfrak g^\vee$.  For
brevity, we suppose that $G$ and $G^\vee$ are simple.  Then $\Theta$
and $\Theta^\vee$ are one-dimensional and $({\mathrm
{Sym}}^{d_i}(\mathfrak g))^G=\Theta^{d_i}\otimes
({\mathrm{Sym}}^{d_i}(\mathfrak g^\vee))^{G^\vee}$.  Hence, if
$\gamma^\vee\in\Theta^\vee$ is picked, we get a correspondence
\begin{equation}\label{lolly}P_i\leftrightarrow (\gamma^\vee)^{d_i}
P_i^\vee   \end{equation} between homogeneous polynomials $P_i\in
({\mathrm {Sym}}^{d_i}(\mathfrak g))^G$ and
$P_i^\vee\in({\mathrm{Sym}}^{d_i}(\mathfrak g^\vee))^{G^\vee}$.

The main reason that these considerations are relevant to gauge
theory is that an invariant quadratic form appears in defining the
Lagrangian.  For example, the kinetic energy of the gauge fields is
commonly written
\begin{equation}\label{kine}-\frac{1}{2e^2}\int_M\Tr\,F\wedge\star
F,\end{equation} where $-\Tr$ is usually regarded as an  invariant
quadratic form on $\mathfrak g$ that is defined {\it a priori}, and
$1/e^2$ is a real number. However, as the theory depends not
separately on the quadratic form $-\Tr$ and the real number $1/e^2$
but only on their product, we may as well combine them
to\footnote{The factor of $4\pi$ here is convenient.  Actually, a
more complete description involves the gauge theory $\theta$ angle
as well.  Then the theory really depends on a complex-valued
invariant quadratic form $\tau=(\theta/2\pi+4\pi i/e^2)(-{\mathrm
{Tr}})$, whose imaginary part is positive definite. For our purposes
here, we omit $\theta$ and set $\gamma={\mathrm {Im}}\,\tau$.}
$\gamma=-(4\pi/e^2)\,\Tr$ and say that the theory simply depends on
an arbitrary choice of a positive definite invariant quadratic form
on $\mathfrak g$.  The $G^\vee$ theory similarly depends on a
quadratic form $\gamma^\vee=-(4\pi/{e^{\vee}}^2)\,\mathrm{Tr}$.  The
relation between the two that follows from electric-magnetic duality
is
\begin{equation}\gamma^\vee=\frac{1}{n_{\frak g}}\gamma^{-1},\end{equation}
where $n_{\frak g}$ is the ratio of length squared of long and short
roots of $G$ or $G^\vee$.

\subsection{Remaining Steps}\label{othersteps}

The program that was started in section \ref{universal} has two
remaining steps: to find the $\B$-model duals of the operators
$\P_i^{(r)}$ of the $\A$-model, and to determine their action on the
space $\mathcal H$ of physical states.

The $\B$-model of $G^\vee$ has a complex adjoint-valued scalar field
$\sigma^\vee$ whose role is somewhat similar to that of $\sigma$ in
the $\A$-model.  We have already encountered this field in eqn.
(\ref{donkey}).

For $G=G^\vee=U(1)$, the action of electric-magnetic duality on
these fields is very simple: $\sigma$ maps to a multiple of
$\sigma^\vee$. For nonabelian $G$ and $G^\vee$, the relation cannot
be as simple as that, since $\sigma$ and $\sigma^\vee$ take values
in different spaces -- they are valued in the complexified Lie
algebras of $G$ and $G^\vee$, respectively. However, $G$-invariant
polynomials in $\sigma$ do transform into $G^\vee$-invariant
polynomials in $\sigma^\vee$ in a way that one would guess from eqn.
(\ref{lolly}):
\begin{equation}\label{pollyx}P_i(\sigma)=(
\sqrt{n_{\mathfrak g}}\gamma^{\vee})^{d_i}P_i^\vee(\sigma^\vee).\end{equation}
So $\P_i=P_i(\sigma)$ maps to a multiple of
$\P_i^\vee=P_i^\vee(\sigma^\vee)$.

We can apply this right away to our familiar example of quantization
on $W=S^2\times I$ with mixed Dirichlet and Neumann boundary
conditions in the $\A$-model, and the corresponding dual boundary
conditions in the $\B$-model.  Picking a point $z\in W$ (or really
$z\in M=W\times\R$), the operator $\P_i(z)=P_i(\sigma(z))$
corresponds in general to a natural cohomology class of degree
$2d_i$ on the $\A$-model moduli space.  However, for the specific
case of $W=S^2\times I$ with our chosen boundary conditions,
 $\P_i(z)$ vanishes, as we explained at the end of
section \ref{universal}.  To see the equivalent vanishing in the
$\B$-model on $W$, we note that $\sigma^\vee$, being the raising
operator of a principal $SL_2$ subgroup of $G^\vee$, is nilpotent.
Hence $P_i^\vee(\sigma^\vee)$ vanishes for every invariant
polynomial $P_i^\vee$.  This is the dual of the vanishing seen in
the $\A$-model.

It is probably more interesting to understand the $\B$-model duals
of those $\A$-model operators that are nonvanishing. For this, we
must understand the duals of the other operators $\P_i^{(r)}$
introduced in section \ref{universal}.  As these operators were
obtained by a descent procedure starting with $\P_i^{(0)}=\P_i$,  we
can find their duals by applying the descent procedure starting with
$\P_i^{\vee(0)}=\P_i^\vee$.  In other words, we look for a family of
$r$-form valued operators $\P_i^{\vee(r)}$, $r=0,\dots,4$, with
$\P_i^{\vee(0)}=\P_i^\vee$ and such that
$(\d+\{Q,~\cdot~\})\widehat\P^\vee_i=0$, where
$\widehat\P^\vee_i=\P_i^{\vee(0)}+\P_i^{\vee(1)}+\dots+\P_i^{\vee(4)}$.
The $\P_i^{\vee(r)}$ are uniquely determined by those conditions and
must be the duals of the $\P_i^{(r)}$.

For our application, the important case is $r=2$.  The explicit
formula for $\P_i^{\vee(2)}$ is very similar to the formula
(\ref{turnsout}) for $\P_i^{(2)}$, except that the curvature $F$
must be replaced by $\star F$:
\begin{equation}\label{burnsout}\P_i^{(2)}=\biggl\langle\frac{\partial
P_i}{\partial\sigma^\vee},\star
F\biggr\rangle+\biggl\langle\frac{\partial^2
P_i}{\partial\sigma^{\vee\,2}},\psi\wedge\psi\biggr\rangle.\end{equation}
Of course, $F$ and $\psi$ are now fields in $G^\vee$ rather than $G$
gauge theory, though we do not indicate this in the notation.

We can now identify the $\B$-model dual of the classes $x_i\in
H^{2d_i-2}(\bN(\rho))$ that were defined in (\ref{gutr}).   We
simply replace $\P_i^{(2)}$ by $(\sqrt{n_{\frak
g}}\gamma^\vee)^{d_i}\P_i^{\vee(2)}$ in the definition of these
classes (the power of $\sqrt{n_{\frak g}}\gamma^\vee$ is from
(\ref{pollyx})), so the dual formula is
\begin{equation}\label{ladr}x_i(S)=(\sqrt{n_{\frak g}}\gamma^\vee)^{d_i}\int_S\P_i^{\vee(2)}.\end{equation}
As in section \ref{universal}, $S$ is a small two-sphere that links
the marked point $w=c\times r\in W$.  We recall that in the
$\B$-model, an external charge in the representation $R^\vee$ is
present at the point $w$.

All we have to do, then, is to evaluate the integral on the right
hand side of (\ref{ladr}).  Since $S$ is a small two-cycle around
the point $w=c\times r$, a nonzero integral can arises only if the
two-form $\P_i^{\vee( 2)}$ has a singularity at $w$.  The reason that
there is such a singularity is that the external charge in the representation
$R^\vee$ produces an electric field, or in other words a
contribution to $\star F$.  In keeping with Coulomb's law, the
electric field is proportional to the inverse of the square of the
distance from the location $w$ of the external charge.  As a result,
$\star F$ has a nonzero integral over $S$.  The electric field due
to the external point charge is proportional to $e^{\vee 2}$ or in
other words to $(\gamma^\vee)^{-1}$.  It is also proportional to the
charge generators, that is, to the matrices that represent the
$G^\vee$ action on $R^\vee$.  Taking this into account, we find that
\begin{equation}\label{zelf}\int_S \P_i^{\vee(2)}=(\gamma^\vee)^{-1}\frac{\partial
P_i^\vee}{\partial\sigma^\vee}.\end{equation} To understand this
formula, observe that as $P_i$ is an invariant polynomial on $\frak
g^\vee$, its derivative $\partial P_i/\partial\sigma^\vee$ can be
understood as an element of the dual space $(\frak g^\vee)^*$;
understanding $(\gamma^\vee)^{-1}$ as a map from $(\frak g^\vee)^*$
to $\frak g^\vee$, the right hand side of (\ref{zelf}) is an element
of $\frak g^\vee$, or in other words an operator that acts on the
space $\mathcal H=R^\vee$ of physical states.

So at last, the  $\A$-model cohomology class $x_i=\int_S P_i^{(2)}$
can be written in the $\B$-model as
\begin{equation}\label{polly}x_i=n_{\frak g}^{d_i/2}(\gamma^\vee)^{d_i-1}\frac{\partial
P_i^\vee(\sigma^\vee)}{\partial\sigma^\vee}.\end{equation}

An illuminating special case of this result is the case that we pick
$P_i$ to be of degree 2, corresponding to an invariant quadratic
form on $\frak g$ and to a two-dimensional class $x\in
H^2(\bN(\rho))$.  In this case, $\partial
P^\vee/\partial\sigma^\vee$ is a Lie algebra element that is linear
in $\sigma^\vee$, and is in fact simply a multiple of $\sigma^\vee$.
In the relevant solution of Nahm's equations, $\sigma^\vee$ is the
raising operator of a principal $SL_2$.  So in other words, the
class  $x\in H^2(\bN(\rho))$ maps to
the raising operator of a principal $SL_2$, acting on $R^\vee$. This
is  a typical fact described in section \ref{chg}.

Finally, we can understand in what sense this result is independent
of the choice of $\gamma^\vee$ (which should be irrelevant in the
$\B$-model). The raising operator of a principal $SL_2$ is
well-defined only up to a scalar multiple. As the right hand side of
eqn. (\ref{polly}) is homogeneous in $\sigma^\vee$ of degree $d_i-1$,
a change in $\gamma^\vee$ can be absorbed in a rescaling of
$\sigma^\vee$; the same rescaling works for all $i$.

\subsection{Compatibility With Fusion}\label{fusion}

\begin{figure}
  \begin{center}
    \includegraphics[width=4in]{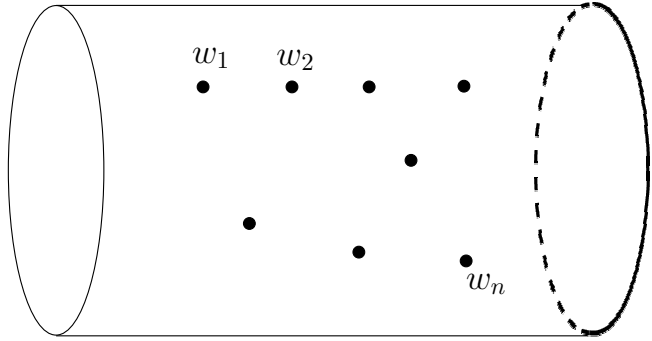}
  \end{center}
\caption{$S^2\times I$ with $n$ marked points (only a few of which
have been labeled) at which 't Hooft or Wilson operators have been
inserted. }
  \label{zocto}\end{figure}

For simplicity, we have considered the case of a single marked point
$w\in W=S^2\times I$.  However, there is an immediate generalization
to the case of several distinct marked points $w_\alpha\in W$,
labeled by representations $R_\alpha^\vee$ of $G^\vee$.  At these
points there is an 't Hooft singularity in the $\A$-model, or an
external charge in the given representation in the $\B$-model (see
fig. \ref{zocto}).

On the $\A$-model side, the moduli space with our usual mixed
boundary conditions is $\mathcal M=\prod_\alpha\bN(\rho_\alpha)$,
where $\rho_\alpha$ is related to $R_\alpha^\vee$ as described in
section \ref{dualg}. This follows from the relation of the Bogomolny
equations to Hecke modifications.  The space of physical states is
the cohomology of $\M$ or
\begin{equation}\label{turngo}\mathcal H=\otimes_\alpha
H^*(\bN(\rho_\alpha)).\end{equation}

On the $\B$-model side, since the solution of Nahm's equations is
unique and irreducible, with a regular pole at one end, the physical
Hilbert space is simply the tensor product of the representations
$R_\alpha^\vee$ associated with the marked points:
\begin{equation}\label{urngo}\mathcal
H=\otimes_\alpha R_\alpha^\vee.\end{equation} The duality map
between (\ref{turngo}) and (\ref{urngo}) is simply induced from the
individual isomorphisms $H^*(\bN(\rho_\alpha))\leftrightarrow
R_\alpha^\vee$.

As discussed in section \ref{conv}, we can also let some of the
points $w_\alpha$ coalesce.  This leads to an operator product
expansion of 't Hooft operators in the $\A$-model, or of Wilson
operators in the $\B$-model.  On the $\B$-model side, the operator
product expansion for Wilson operators corresponds to the classical
tensor product $R^\vee_\alpha\otimes R^\vee_\beta=\oplus_\gamma
N^\gamma_{\alpha\beta}R^\vee_\gamma$.  The corresponding $\A$-model
picture is more complicated and is described in gauge theory terms
in section 10.4 of \cite{KW}.

The only observation that we will add here is that the operator
product expansion for Wilson or 't Hooft operators commutes with the
action of the group $\mathcal T$ described at the end of section
\ref{chg}. We recall that $\mathcal T$ is generated on the
$\A$-model side by the grading of the cohomology by degree and the
action of the cohomology classes $x_i(S)$.  For example, consider
the grading of the $\A$-model cohomology by  degree. With
$\M=\prod_\alpha \bN(\rho_\alpha)$, the operator that grades
$\mathcal H=H^*(\M)$ by the degree of a cohomology class is the sum
of the corresponding operators on the individual factors
$H^*(\bN(\rho_\alpha))$. The operator product expansion of 't Hooft
operators commutes with the degree or ghost number symmetry, which
after all originates as a symmetry group (a group of $R$-symmetries)
of the full $\NN=4$ super Yang-Mills theory. So after fusing some of
the 't Hooft operators together, the action of the ghost number
symmetry is unchanged. Similarly, the dual $\B$-model grading is by
the generator $t_3$ of a maximal torus of a principal $SL_2$
subgroup of $G^\vee$. Again, the linear transformation by which
$t_3$ acts on $\mathcal H=\otimes_\alpha R_\alpha^\vee$ is the sum
of the corresponding linear transformations for the individual
$R_\alpha^\vee$. This linear transformation is unchanged if some of the
points are fused together, since it originates as a combination of
an $R$-symmetry and a gauge transformation, both of which are
symmetries of the full theory and therefore of the operator product
expansion of Wilson operators.

\begin{figure}
  \begin{center}
    \includegraphics[width=3in]{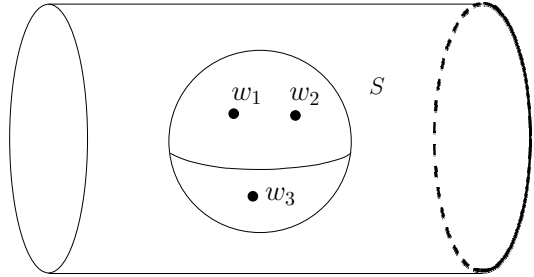}
  \end{center}
\caption{A two-sphere $S$ (at fixed time) surrounding all of the
marked points $w_\alpha\in S^2\times I$.  (In this example, there
are three marked points).  $S$ is homologous to a sum of two-spheres
$S_\alpha$, each of them linking just one of the $w_\alpha$.}
  \label{pocto}\end{figure}

A similar story holds for the linear transformations that correspond
to the gauge theory cohomology classes $x_i$ introduced in section
\ref{chg}.  Let $S$ be a two-cycle that encloses all of the marked
points $w_\alpha$, as indicated in fig. \ref{pocto}.  In the
$\A$-model, we have   $x_i=\int_S \P_i^{(2)}$, while in the
$\B$-model the analog is $x_i=\int_S\P_i^{\vee(2)}$.  These
definitions make it clear that nothing happens to $x_i$ if we fuse
together some of the points $w_\alpha$ that are contained inside
$S$.

\begin{figure}
  \begin{center}
    \includegraphics[width=3in]{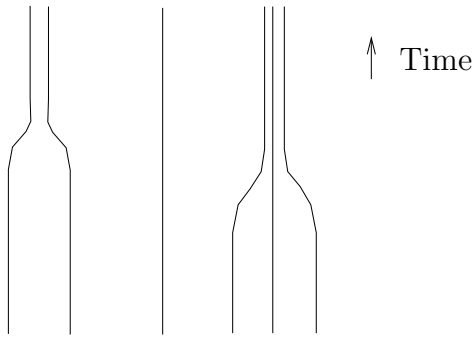}
  \end{center}
\caption{The operator product expansion as a time-dependent process.
Time runs vertically in the figure.  In the past, there are $n$
distinct marked points with insertions of Wilson or 't Hooft
operators. In the future, the points fuse together in various
groups. (Complete fusion is not shown in the figure.) In this
example, $n=6$ and the groups are of sizes 2, 1, and 3. If as in
fig. \ref{pocto} we add a two-sphere $S$ that surrounds all of the
points, then by topological invariance we could move it to the past
where it acts on $n$ distinct line operators or to the future where
it acts on a smaller number of line operators created by fusion.
Hence the action of the operators $x_i(S)$ commutes with the
operator product expansion of Wilson or 't Hooft operators.}
  \label{rocto}\end{figure}
It is illuminating here to think of the fusion as a time-dependent
process. We go back to a four-dimensional picture on $M=W\times \R$,
where $\R$ parametrizes the time, and instead of thinking of the
marked points as having time-independent positions (as we have done
so far in this article), we take them to be separate in the past and
to possibly fuse together (in arbitrary subsets) in the far future,
as in fig. \ref{rocto}. The surface $S$ is located at a fixed time,
but topological invariance means that we can place it in the far
past -- acting on the Hilbert space of a collection of isolated
points $w_\alpha$ -- or in the far future -- after some fusing may
have occurred.  So fusion commutes\footnote{The ability to move the
$x_i(S)$ backwards or forwards in time also means that they are
central -- they commute with any other operators that may act on the
Wlson or 't Hooft operators.}  with the action of $x_i(S)$.

Finally, we want to see that for each $i$, the linear transformation
by which $x_i$ acts on the physical Hilbert space $\mathcal
H=\otimes_\alpha H^*(\bN(\rho_\alpha))=\otimes_\alpha R_\alpha^\vee$
can be written as a sum of the linear transformations by which $x_i$
would act on the individual factors $H^*(\bN(\rho_\alpha))$ or
$R_\alpha^\vee$. For each $\alpha$, let $S_\alpha$ be a two-cycle
that encloses only the single marked point $w_\alpha$. Then $S$ is
homologous to the sum of the $S_\alpha$. Hence $\int_S
\P_i^{(2)}=\sum_\alpha\int_{S_\alpha}\P_i^{(2)}$, and similarly
$\int_S\P_i^{\vee(2)}=\sum_\alpha\int_{S_\alpha}\P_i^{\vee(2)}$. So
in either the $\A$-model or the $\B$-model, $x_i$ acts on $\mathcal
H$ by the sum of the linear transformations by which $x_i$ would act
on a single factor $H^*(\bN(\rho_\alpha))$ or $R_\alpha^\vee$.

\section{From Physical Yang-Mills Theory To Topological Field Theory}\label{details}

In sections \ref{twisting} and \ref{scalars}, we will describe  some
details of the relation between supersymmetric Yang-Mills theory and
topological field theory in four dimensions that were omitted in
section \ref{gaugetheory}.

In section \ref{reduction}, we discuss the compactification (not
reduction) of the theory to three dimensions, hopefully shedding
light on some recent mathematical work \cite{BN}.

In section \ref{realdetails}, we explain the claim of section
\ref{subtlety} that with the boundary conditions that we chose on
$W=S^2\times I$, Nahm's equations have a unique solution.  Finally,
in section \ref{genker}, we explain the relation of the dual of
Dirichlet boundary conditions to the universal kernel of geometric
Langlands.

\subsection{Twisting}\label{twisting}

\def\Y{\mathcal Y}

We begin by reviewing the ``twisting'' procedure by which
topological field theories can be constructed starting from
supersymmetric Yang-Mills theory in four dimensions. The original
example involved starting with $\NN=2$ super Yang-Mills theory; the
twisted theory is then essentially unique and is related to
Donaldson theory \cite{EWT}.  Starting from $\NN=4$ super Yang-Mills
theory, there are three choices \cite{VW}, one of which is related
to geometric Langlands \cite{KW}.

We begin by considering  $\NN=4$ super Yang-Mills theory on
Euclidean space $\R^4$. The rotation group is $SO(4)$.  We denote
the positive and negative spin representations of the double cover
$\Spin(4)$  as $V_+$ and $V_-$, respectively; they are both
two-dimensional.  One important point is that although $\NN=4$ super
Yang-Mills theory is conformally invariant both classically and
quantum mechanically, the twisting procedure does not use this
conformal invariance.  (A close analogy of the construction with the
twisting of $\NN=2$ super Yang-Mills theory would not be possible if
we had to make use of conformal invariance, since $\NN=2$ super
Yang-Mills theory is not conformally invariant quantum
mechanically.)

$\NN=4$ super Yang-Mills theory also has an $R$-symmetry group
$\Spin(6)$. An $R$-symmetry group is simply a group of symmetries
that acts by automorphisms of the supersymmetries, while acting
trivially on spacetime.  The group $\Spin(6)$ has positive and
negative spin representations that we will call $U_+$ and $U_-$.
They are both of dimension 4. The supersymmetries of $\NN=4$ super
Yang-Mills theory transform under $\Spin(4)\times \Spin(6)$ as
\begin{equation}\label{trol}\Y=V_+\otimes U_+\oplus V_-\otimes U_-.\end{equation}

Classically \cite{BSS}, it is possible to construct $\NN=4$
super-Yang Mills theory by dimensional reduction from ten
dimensions, that is from $\R^{10}$. This entails an embedding
$(\Spin(4)\times \Spin(6))/\Z_2\subset\Spin(10)$.  In this way of
constructing the $\NN=4$ theory,  $\Y$ simply corresponds to one of
the irreducible spin representations of $\Spin(10)$.  The
supersymmetry algebra in ten dimensions reads
\begin{equation}\label{zelk}\{Q_\gamma,Q_\delta\}=\sum_{I=1}^{10}
\Gamma^I_{\gamma\delta}P_I,\end{equation} where the notation is as
follows.  $Q_\gamma$ and $Q_\delta$ are two supersymmetry charges,
corresponding to elements of $\Y$.  The $P_I$ generate the
translation symmetries of $\R^{10}$.  And the $\Gamma_I$ are the
generators of the Clifford algebra, understood as bilinear maps
${\mathrm {Sym}}^2\,\Y\to V_{10}$, where $V_{10}$ is the
ten-dimensional representation of $\Spin(10)$. Reduction to four
dimensions is achieved by requiring the fields to be independent of
the last six coordinates of $\R^{10}$.  This reduces $\Spin(10)$
symmetry to the subgroup $(\Spin(4)\times \Spin(6))/\Z_2$ considered
in the last paragraph.  In the reduced theory,\footnote{It is
possible \cite{OW} to pick boundary conditions such that the $P_I$,
$I\geq 5$, survive in the reduced theory as central charges
(electric charges) that commute with all local operators (and in
this case magnetic charges appear in the algebra as additional
central charges). We will not be interested here in such boundary
conditions.  In any event, the automorphism of the algebra of local
operators generated by the $P_I$  always vanishes; this is what we
need in the following arguments. } the $P_I$ vanish in eqn.
(\ref{zelk}) for $5\leq I\leq 10$. Therefore, in the reduced theory,
the right hand side of (\ref{zelk}) contains precisely the four
operators $P_I$, $I=1,\dots,4$.

\remark\label{pok} In particular, in the theory reduced to four dimensions,
there is no $\Spin(4)$-invariant
operator on the right hand side of (\ref{zelk}).  On the other hand,
 the right hand side of (\ref{zelk}) is
$\Spin(6)$-invariant.
\medskip

The idea of twisting is to replace $\Spin(4)$ by another subgroup of
$(\Spin(4)\times \Spin(6))/\Z_2$ that acts in the same way on
spacetime, but has some convenient properties that will be
described. This is accomplished by picking a homomorphism
$\lambda:\Spin(4)\to \Spin(6)$.    Then we extend this to an
embedding $(1\times\lambda):\Spin(4)\to \Spin(4)\times \Spin(6)$ and
we define $\Spin'(4)= (1\times\lambda)(\Spin(4))$. The twisted
theory is one in which the ordinary rotation group $\Spin(4)$ is
replaced by $\Spin'(4)$.  In other words, whenever we make a
rotation of $\R^4$ by an element $f\in\Spin(4)$, we accompany this
by a $\Spin(6)$ transformation $\lambda(f)$.

We want to pick $\lambda$ so that the $\Spin(4)\times \Spin(6)$
module $\Y$  contains a nonzero $\Spin'(4)$ invariant. Supposing
that this is the case, pick such an invariant and write $ Q$ for the
corresponding supersymmetry. $Q$ automatically satisfies the
fundamental condition $Q^2=0$.  The reason for this is that $Q^2$ is
$\Spin'(4)$ invariant and (since $Q$ is a linear combination of the
$Q_\gamma$) can be computed from (\ref{zelk}).  But in view of
Remark \ref{pok}, this is no $\Spin'(4)$ invariant on the right hand
side of (\ref{zelk}).

Since $Q^2=0$, one can pass from $\NN=4$ super Yang-Mills theory to
a much ``smaller'' theory by taking the cohomology of $Q$.  One
considers only operators (or states) that commute with $Q$ (or are
annihilated by $Q$) modulo operators of the form $\{Q,\dots\}$ (or
states in the image of $Q$).

It is possible to state a simple condition under which the small
theory can be extended to a topological field theory. The condition
is that  the stress tensor $T$ of the theory, which measures the
response of the theory to a change in the metric of $\R^4$, must be
trivial in the cohomology of $Q$, that is it must be of the form
$T=\{Q,\Lambda\}$ for some $\Lambda$.  In practice, this condition
is always satisfied in four dimensions. Given this, one can promote
the ``local'' construction on $\R^4$ sketched in the last few
paragraphs to a ``global'' construction that makes sense on a rather
general smooth four-manifold $M$. (Depending on $\lambda$, $M$ may
require some additional structure such as an orientation or a spin
structure; however, for the choice of $\lambda$ that leads to
geometric Langlands, no such additional structure is required.)

Three possible twists of $\NN=4$ super Yang-Mills theory lead to
topological field theories.  Two of these are close cousins of
Donaldson theory, and the third is related to geometric Langlands.

The twist that leads to geometric Langlands is easily described.
$SO(6)=\Spin(6)/\Z_2$ has a obvious $SO(4)\times SO(2)$ subgroup
\begin{equation}\label{blocks}\begin{pmatrix}*&*&*&*&0&0\\ *&*&*&*&0&0\\ *&*&*&*&0&0\\ *&*&*&*&0&0\\
           0&0&0&0&*&*\\   0&0&0&0&*&* \end{pmatrix}.\end{equation}
           Taking the double cover, $\Spin(6)$ has commuting $\Spin(4)$ and $\Spin(2)$ subgroups whose
           centers coincide,
            and hence a global embedding
            \begin{equation}\label{pony}(\Spin(4)\times \Spin(2))/\Z_2\subset \Spin(6).\end{equation}
We simply take $\lambda:\Spin(4)\to\Spin(6)$ to be an isomorphism onto this $\Spin(4)$ subgroup of
$\Spin(6)$.
Since $\Spin(2)$ commutes with the image of $\lambda$, it becomes a global symmetry of the model.
This actually is the
group $\Spin(2)$ that played an important role in section \ref{space}.

The spin representations $U_\pm$ of $\Spin(6)$ decompose under $\Spin(4)\times \Spin(2)$ as
\begin{align}\label{delk} \notag   U_+ & = V_+^1\oplus V_-^{-1} \\
                                                            U_-& = V_+^{-1}\oplus V_-^1.\end{align}
Here the notation is as follows.  As before, $V_+$ and $V_-$ are the two spin representations of
$\Spin(4)$.
As for $\Spin(2)$, it is abelian and isomorphic to $U(1)$. Its  spin representations are
one-dimensional representations
of $U(1)$ of ``charge'' 1 and $-1$; the charge  is indicated by the superscripts $\pm 1$ in
eqn. (\ref{delk}).

Now in view of eqn. (\ref{trol}), the supersymmetries of the theory transform under
$\Spin'(4)\times \Spin(2)$ as
\begin{equation}\label{helk} V_+\otimes (V_+^1\oplus V_-^{-1})\oplus V_-\otimes
(V_+^{-1}\oplus V_-^{1}).\end{equation}
We want to find the $\Spin'(4)$ invariants.  Decomposing (\ref{helk}) into a
direct sum of irreducibles, both $V_+\otimes
V_+^1$ and $V_-\otimes V_-^1$ contain a one-dimensional $\Spin'(4)$-invariant
subspace, while there are
no invariants in $V_\pm\otimes V_\mp^{-1}$.

Let us write $Q_+$ and $Q_-$ for $\Spin'(4)$-invariant
supersymmetries derived from the invariant part of $V_+\otimes
V_+^1$ and $V_-\otimes V_-^1$, respectively.  Notice that they both
transform under $\Spin(2)$ with charge 1.  A general complex linear
combination
\begin{equation}\label{mork}Q=uQ_++vQ_-\end{equation} is
$\Spin'(4)$-invariant and also has charge 1. It turns out also that
any such $Q$ (with $u$ and $v$ not both zero) obeys the condition
for defining a topological field theory -- the stress tensor can be
written as $T=\{Q,\dots\}$. The topological field theory that we get
by passing to the cohomology of $Q$ is invariant under rescaling $Q$
by a nonzero complex number. So we should think of $u$ and $v$ as
homogeneous coordinates on a copy of $\CP^1$ that parametrizes a
family of topological field theories.

Because $\Spin(2)\cong U(1)$, its representations are labeled by
integers, corresponding to the characters $\exp(i\theta)\to
\exp(in\theta)$, $n\in \Z$.   The action of $\Spin(2)$ gives a
$\Z$-grading of the full physical Hilbert space $\widehat {\mathcal
H}$ of $\NN=4$ super Yang-Mills theory.

In the topological field theory, we want a $\Z$-grading not of
$\widehat{\mathcal H}$, but of a vastly smaller space $\mathcal H$
-- the cohomology of $Q$. In order for the cohomology of $Q$ to be
$\Z$-graded, we require that $Q$ should transform in a definite
character of $\Spin(2)$. This is true, for any choice of $u$ and
$v$, because
 both $Q_+$ and $Q_-$ transform with the same character of $\Spin(2)$ -- what we have called
charge 1.  So any complex linear combination $Q=u Q_++v Q_-$ also
has charge or degree 1, and the cohomology of $Q$ is $\Z$-graded.

If it were the case, for example, that $Q_+$ and $Q_-$ had
respectively charge $1$ and $-1$, then a generic complex linear
combination $Q=uQ_++vQ_-$ would not have definite charge, and its
cohomology would be only $\Z_2$-graded.  In section \ref{reduction},
we will describe a situation in which something similar to that
occurs.

\subsubsection{A Slight Complication}\label{complication}

Roughly speaking, the $\A$-model and the $\B$-model correspond to
different values of the ratio $v/u$.  The full details are a little
more complicated, and involve also the coupling parameter
$\tau=\theta/2\pi+4\pi i/e^2$ of the gauge theory, as explained in
\cite{KW}, section 3.5.

The complication arises because the
Lagrangian of the theory cannot be written in the form
$\{Q,\dots\}$, but is of this form only modulo a multiple of the
topological invariant $\int_M\Tr\,F\wedge F$.  Consequently, the
topological field theory depends not only on the twisting parameter
$v/u$, but also on $\tau$.  Actually, the topological field theory
depends on $\tau$ and the twisting parameter only via a single
parameter defined in eqn. (3.50) of \cite{KW}; as a result, it is
true that twisting leads to a family of topological field theories
parametrized by $\CP^1$ and that the $\A$-model and the $\B$-model
correspond to two points in this space.

The expression $\int_M\Tr\,F\wedge F$ actually has another
interpretation.  Let $P(\sigma)$ be an invariant quadratic
polynomial on the Lie algebra $\frak g$.  Applying to $P$ the
construction of section \ref{universal}, we construct a sequence of
$r$-form valued operators $\P^{(r)}$, $r=0,\dots,4$ with
$\P^{(0)}=P(\sigma)$ and $(\d+\{Q,\cdot\})\sum_r\P^{(r)}=0$.  If we
pick $P$ correctly, then $\P^{(4)}=\frac{1}{8\pi^2}\Tr\,F\wedge F$.
The integral $\int_M\P^{(4)}=\frac{1}{8\pi^2}\int_M\Tr\,F\wedge F$
(which is none other than the instanton number) is $Q$-invariant but
is nontrivial in the cohomology of $Q$. It is this fact that causes
the coupling parameter $\tau$ to be relevant in the topological
field theory.

In a superficially similar situation that will be considered in
section \ref{reduction}, $\int_M \P^{(4)}$ will disappear from the
$Q$-cohomology (by ``canceling'' a certain integral of $\P^{(3)}$,
which will also disappear at the same time).  This being so, $\tau$
will be irrelevant in the topological field theory, which will
depend only on the choice of $Q$.

\subsection{Scalar Fields In  The Twisted Theory}\label{scalars}

Now we want to describe the bosonic fields of $\NN=4$ super
Yang-Mills theory, before and after twisting.  In ten dimensions,
the only bosonic field is the connection $\hat A$.  Writing $\hat
A=\sum_{I=1}^4A_I\d x^I +\sum_{J=1}^6 A_J \d x^J$, we can
parametrize $\hat A$ by the four-dimensional connection
$A=\sum_{I=1}^4 A_I\,\d x^I$ and six scalar fields $\Phi_I=A_{4+I}$,
$I=1,\dots,6$ that are valued in the adjoint representation of the
gauge group $G$.

In particular, the six scalar fields $\Phi$ transform in the
``vector'' representation of $SO(6)=\Spin(6)/\Z_2$. Under the
embedding $SO(4)\times SO(2) \subset SO(6)$ sketched in eqn.
(\ref{blocks}), $\Phi$  splits into ``upper'' components that
transform under $SO(4)$ and ``lower'' components that transform
under $SO(2)$.

In the twisted theory on a general four-manifold $M$, the upper
components are interpreted as an $\mathrm{ad}(E)$-valued one-form
$\phi$.     Twisting  transforms $\phi$ from a collection of four
scalar fields (or zero-forms) into a one-form.  In other words, the
upper components of $\phi$ are invariant under $\Spin(4)$, but
transform under  $\Spin'(4)$ in such a way that it is natural to
interpret $\phi=\sum_{I=1}^4\Phi_I\d x^I$ as a one-form. This
one-form entered prominently in section \ref{gaugetheory}. In the
$\B$-model, it combines with $A$ to the complex connection
$\AA=A+i\phi$, and in the $\A$-model, it appears with $A$ in the
elliptic differential equations $F-\phi\wedge\phi=\star \d_A\phi$.

The lower components of $\Phi$ are a pair of $\mathrm{ad}(E)$-valued
scalar fields that transform trivially under $\Spin'(4)$ but in a
real two-dimensional representation of  $\Spin(2)$. In section
\ref{backtogvee}, these fields were called $X_1$ and $X_2$ and were
combined into a complex field $\sigma=(X_1+iX_2)/\sqrt 2$.  The
field $\sigma$ has charge or degree 2, for the following reason.  We
defined the charge so that the fundamental representation of
$\Spin(2)$ has charge 1, so the fundamental representation of
$SO(2)=\Spin(2)/\Z_2$ has charge 2.  The fields $X_1$ and $X_2$
transform in the fundamental representation of $SO(2)$, as is clear
from the embedding $SO(4)\times SO(2)\subset SO(6)$. In the
$\A$-model, $\sigma$ can be viewed as part of the Cartan model of
the equivariant cohomology of the gauge group acting on the fields
$(A,\phi)$.  In the $\B$-model, its role was described in section
\ref{backtogvee}.

All of this holds  on a generic four-manifold $M$.  However, matters
simplify if $M$ is the product of a three-manifold $M_3$ with a
one-manifold $M_1$. Here $M_1$ may be either $\R$ or $S^1$ or  a
compact interval $I$ with some boundary conditions chosen.
Topological field theory on $M$ does not really depend on what
metric is chosen on $M$, but if $M$ is a product, it is simplest to
do the computations with a product metric. The cotangent bundle of
$M$ then splits metrically (as well as topologically) as a direct
sum $T^*M=T^*M_3\oplus T^*M_1$, where the connection on $T^*M_1$ is
trivial.

We now should reexamine the four ``upper'' components of $\Phi$ that
for generic $M$ are interpreted after twisting as a one-form $\phi$.
In the case of the $3\oplus 1$ split of the last paragraph, only
three components of $\Phi$ are twisted.  They can be interpreted as
a one-form on $M_3$.    As for the fourth ``upper'' component, it is
a one-form on $M_1$, but the cotangent bundle of $M_1$ is completely
trivial -- topologically, metrically, and from the point of view of
the Riemannian connection. So in this particular situation, twisting
has done nothing at all to this scalar field.   Since it is
unaffected by the twisting, just like the ``lower'' components $X_1$
and $X_2$, we may as well combine it with them and call it $X_3$.

The $\Spin(2)$ global symmetry of $\NN=4$ twisted super Yang-Mills
theory on a generic $M$ is now promoted to $\Spin(3)$, rotating
$X_1,X_2,X_3$.  This is the $\Spin(3)$ symmetry that mysteriously
appeared when we derived Nahm's equations in section
\ref{backtogvee}.     The example in that section  and in most of
section \ref{gaugetheory} was $M=\R\times S^2\times I$, which can be
decomposed as $M_3\times M_1$ in more than one way. The
decomposition that is relevant for understanding section
\ref{backtogvee} is $M_3=\R\times S^2$, $M_1=I$. Indeed, the formula
$X_3=\phi_y$ of eqn. (\ref{egg}) shows that $X_3$ is the component
of $\phi$ in the $I$ direction.

Though physical Yang-Mills theory on $M_3\times M_1$ (after twisting
but before passing to the $Q$ cohomology) has $\Spin(3)$ symmetry,
the topological field theory that we get by taking the $Q$
cohomology does not. That is because $Q$ does not transform in a
one-dimensional representation of $\Spin(3)$.  In fact, it lies in a
two-dimensional representation of $\Spin(3)$.

\def\VV{{\mathcal V}}
\subsection{More General Construction In Three Dimensions}\label{reduction}

We will now make a digression aimed at making contact with some
recent mathematical work \cite{BN}. At the end of section
\ref{scalars}, we considered a four-dimensional  topological field
theory specialized to a four-manifold with a product structure.
Henceforth we take this to be specifically $M=M_3\times S^1$.
Keeping $S^1$ fixed and letting $M_3$ vary, the four-dimensional
topological field theory reduces to a three-dimensional one.

It is possible to modify the construction slightly to get from
$\NN=4$ super Yang-Mills in four dimensions a three-dimensional
topological field theory that does {\it not} quite come in this way
from a four-dimensional topological field theory.  Roughly speaking,
to do this, we require $Q$ to have only $\Spin'(3)$ invariance, not
$\Spin'(4)$ invariance.

To explain the construction in more detail, begin with $\NN=4$ super
Yang-Mills theory on $\R^3\times S^1$. The spin group of $\R^3$ is
$\Spin(3)$, and of course the $R$-symmetry group of the theory is
still $\Spin(6)$. Now we want to pick a homomorphism
$\tilde\lambda:\Spin(3)\to\Spin(6)$ and to define $\Spin'(3)$ as the
image of $(1\times\tilde\lambda):\Spin(3)\to \Spin(3)\times
\Spin(6)$.  We simply define $\tilde\lambda$ to be the restriction
to $\Spin(3)$ of the homomorphism $\lambda:\Spin(4)\to\Spin(6)$ that
we used before. To say this differently, we now begin with a
subgroup $(\Spin_1(3)\times \Spin_2(3))/\Z_2\subset \Spin(6)$ (here
$\Spin_i(3)$, $i=1,2$, are two commuting copies of $\Spin(3)$). We
define $\Spin'(3)$ to be the diagonal product of $\Spin(3)\times
\Spin_1(3)\subset \Spin(3)\times \Spin(6)$.

Clearly, $\Spin'(3)$ commutes with the group $F=\Spin_2(3)$, which
is yet another copy of $\Spin(3)$. $F$ will play the role that was
played in sections \ref{twisting} and \ref{scalars} by $\Spin(2)$.
The reason for the extension of $\Spin(2)$ to $\Spin(3)$ is the same
as in section \ref{scalars} -- only three scalar fields have been
twisted, not four. We will also be interested in the
complexification of $F$, which is $F_\C=\Spin(3,\C)\cong SL(2,\C)$.

To construct a three-dimensional topological field theory, we must
pick a $\Spin'(3)$-invariant supercharge.  So let us determine how
the supercharges transform under $\Spin'(3)\times F$.  We write $V$,
$V_1$ and $V_2$ for the spin representations of $\Spin(3)$,
$\Spin_1(3)$ and $\Spin_2(3)$.  The two spin representations $V_\pm$
of $\Spin(4)$ are both equivalent to $V$ when restricted to
$\Spin(3)$.  Similarly, the two spin representations $U_\pm$ of
$\Spin(6)$ are both equivalent under $(\Spin_1(3)\times
\Spin_2(3))/\Z_2$ to $V_1\otimes V_2$. So as a $\Spin(3)\times
\Spin_1(3)\times \Spin_2(3)$ module, the space of supersymmetries is
\begin{equation}\mathcal Y=V\otimes V_1\otimes V_2\otimes \C^2.\end{equation}
We restrict to $\Spin'(3)\times \Spin_2(3)$ by setting $V_1=V$,
giving $\mathcal Y=V\otimes V\otimes V_2\otimes \C^2$ The first step
in constructing three-dimensional supersymmetric field theories is
to extract the $\Spin'(3)$-invariant subspace. The
$\Spin'(3)$-invariant subspace of $V\otimes V$ is one-dimensional,
so the $\Spin'(3)$-invariant subspace  of $\mathcal Y$ is
four-dimensional.  We call this subspace $J$.  As an $F$-module, $J$
is isomorphic to $V_2\otimes\C^2$, where $V_2$ is a two-dimensional
module for $F_\C\cong SL(2,\C)$.

If $Q$ is the supersymmetry corresponding to a generic point in $J$,
it is {\it not} true that $Q^2=0$.  We can see this from
(\ref{zelk}).  Though there is no $\Spin'(4)$-invariant on the right
hand side of (\ref{zelk}), there is an essentially unique
$\Spin'(3)$ invariant.  It is the generator of the rotation of
$S^1$, the second factor of $\R^3\times S^1$.  Let us call this
generator $\VV$.  A generic $\Spin'(3)$-invariant supersymmetry
squares not to zero but to a multiple of $\VV$.  On the $\Spin'(3)$
invariant subspace $J$, (\ref{zelk}) reduces to something that in
coordinates looks like
\begin{equation}\label{pelk}\{Q_\alpha,Q_\beta\}=\delta_{\alpha\beta}\VV.\end{equation}
Intrinsically, $\delta_{\alpha\beta}$ is a quadratic form $(~,~)$ on the
four-dimensional vector space $J$.  This quadratic form is obviously
$F_\C$-invariant, and this is actually enough to ensure that it is
nondegenerate, given that $J\cong V_2\otimes \C^2$.  Indeed, the
quadratic form is the tensor product of an $F_\C$-invariant skew
form on $V_2$ and a nonzero (and therefore nondegenerate) skew form
on $\C^2$.  The skew form on $\C^2$ is invariant under a group
$\tilde F_\C$ that is another copy of $SL(2,\C)$.  $\tilde F_\C$ is therefore a
group of symmetries of the quadratic form, though there is no natural
way to make it act on the states and operators of the full theory.

Suppose that $Q$ is a $\Spin'(3)$-invariant supersymmetry with
$Q^2=\VV$ (or equivalently, $Q^2$ a nonzero multiple of $\VV$). Can
we use $Q$ as a differential to construct a topological field
theory? Superficially the answer is ``no,'' since $Q^2$ is nonzero.
However, $\VV$ generates a symmetry -- a compact group of rotations
of $\R^3\times S^1$ -- and we can restrict to $\VV$-invariant
operators and states.  In this smaller space, $Q^2=0$ and we can
pass to the cohomology of $Q$.  In fact, similar constructions have
been made previously \cite{Nek,P}. These constructions,
respectively, involve non-free $S^1$ actions on $\R^4$ or $S^4$. The
construction we are describing here  is similar but simpler as it
involves a free $S^1$ action.

The relation $Q^2=\VV$ is reminiscent of equivariant cohomology.
Consider a $U(1)$ action on a manifold $B$ generated by a vector
field $V$. Localized equivariant cohomology can be described by the
operator $\d_V=\d+\iota_V$ acting on differential forms on $B $;
here $\iota_V$ is the operator of contraction with $V$.  One has
$\d_V^2=\mathcal L_V$, where $\mathcal L_V$ is the Lie derivative
with respect to $V$.  The operator $\d_V$ was related to
supersymmetric nonlinear sigma models in \cite{EWM} and interpreted
in equivariant cohomology in \cite{ABM}.  In our problem, since
$\VV$ generates the natural $S^1$ action on $M_4=M_3\times S^1$, the
relation $Q^2=\VV$ is suggestive of localized equivariant cohomology
for this action.  This connection is made much more precise in
\cite{Nek,P}.

\def\PP{{\Bbb P}}
Up to scaling by a nonzero complex number, $Q$ corresponds {\it a
priori} to an arbitrary point in the projective space $\PP(J)\cong
\CP^3$. But it is not true that $\CP^3$ parametrizes a family of
inequivalent topological field theories. If $f$ is any invertible
operator acting on the Hilbert space $\mathcal H$ of $\NN=4$ super
Yang-Mills theory, then $Q$ and $fQf^{-1}$ lead to equivalent
topological field theories.  In particular, picking $f\in F_\C\cong
SL(2,\C)$, we see that to classify the three-dimensional topological
field theories that emerge from this construction, we must divide by
the action of $F_\C$ on $\CP^3$.

Let us first classify those topological field theories for which
$Q^2=0$.  These correspond to the zeroes of the  nondegenerate
quadratic form $(~,~)$ on $\PP(J)$.  They form a nondegenerate
quadric $\mathcal Q$, which is a copy of $\CP^1\times \CP^1$.  This
particular copy of $\CP^1\times\CP^1$ is a homogeneous space for the
group $SO(4,\C)\cong (F_\C\times \tilde F_\C)/\Z_2$ that acts on
$\PP(J)$ preserving the quadric, so we write it as
$\CP^1\times\tilde{\CP}{}^1$.  Here $\CP^1$ is a homogeneous space
for $F_\C$, and $\tilde{\CP}{}^1$ is a homogeneous space for $\tilde
F_\C$.  The quotient $(\CP^1\times \tilde{\CP}{}^1)/F_\C$ is just a
copy of $\tilde{\CP}{}^1$. However, $F_\C$ does not act freely on
$\tilde\CP^1$.  Each point in $\tilde\CP^1$ is left fixed by a Borel
subgroup $\mathcal B$ of $F_\C$, isomorphic to
\begin{equation}\label{polky}\begin{pmatrix}* & * \\ 0 & * \end{pmatrix}.\end{equation}
In the topological field theory associated to a particular choice of
$Q$, this Borel group acts as a group of symmetries. In particular,
the cohomology of $Q$ is $\Z$-graded by the action of the diagonal
matrices in $\mathcal B$.

So we have a family of $\Z$-graded three-dimensional topological field theories, parametrized by a copy of $\CP^1$ (at this
point we drop the tilde), with the property
that $Q^2=0$.  Actually, these are simply the examples that come by compactification on $S^1$ of a four-dimensional
topological field theory.

To get something new, we consider the examples for which $Q^2$ is a
nonzero multiple of $\VV$.  Notice that $\PP(J)$ is a complex
manifold of complex dimension 3, as is $F_\C$.  This makes it
possible for the complement of the quadric $\mathcal Q\subset
\PP(J)$ to consist of a single $F_\C$ orbit.  This is in fact the
situation.  Bearing in mind the decomposition $J\cong V_2\otimes
\C^2$, where $F_\C$ acts on the first factor and $\tilde F_\C$ on the
second, we can think of an element of $J$ as a $2\times 2$ matrix
$K_{A\dot A}$, $A,\dot A=1,2$, with $F_\C$ and $\tilde F_\C$ acting
on $K$ respectively on the left and right.  In this representation,
the $F_\C\times \tilde F_\C$-invariant quadratic form is $K\to
\det(K)$ and the condition for $K$ not to be a null vector for the
quadratic form is that it should be an invertible matrix.  But any
two invertible matrices are equivalent under the action of
$F_\C\times \C^*$ ($F_\C$ acts on the $2\times 2$ matrix $K$ by left
multiplication, while $\C^*$ acts by scaling $K\to \lambda K$,
$\lambda\in\C^*$; we must divide by $\C^*$ since we view $K$ as an
element of the projective space $\PP(J)$). So as claimed, the
complement of the quadric in $\PP(J)$ is a single $F_\C$ orbit.

Although the left action of $F_\C$ on the space of invertible
$2\times 2$ matrices is free, when we project to $\PP(J)$, the
action becomes only semi-free (that is, the stabilizer of a point is
a finite group).  In fact, $F_\C\cong SL(2,\C)$ contains a central
subgroup $\Z_2$ consisting of the matrices $-1$ and $1$.  These
matrices act freely on $\PP(J)$ and the subgroup of $F_\C$ that
leaves fixed a point in $\PP(J)$ that is not on the quadric is
$\Z_2$.  So if $Q$ corresponds to a point not on the quadric, then
its cohomology is $\Z_2$-graded, but not $\Z$-graded.

So we can summarize what three-dimensional topological field
theories arise from this construction.  There is the usual $\CP^1$
family of theories that arise by compactification from four
dimensions.  Two points in this family are the $\A$-model and
$\B$-model of $G$ (which are equivalent,  respectively, to the
$\B$-model and $\A$-model of $G^\vee$). The generic point in this
family corresponds to what is sometimes called quantum geometric
Langlands (of $G$ or equivalently of $G^\vee$). There is one more
theory that does {\it not} arise by compactification of a
four-dimensional theory. It is only $\Z_2$ graded and as we explain
momentarily does not distinguish $G$ from $G^\vee$.

What we have established so far is really that by varying $Q$ at a fixed value of the coupling parameter $\tau$ of
the theory, we can construct only one new theory.  In section \ref{vanishing}, we will show that because of vanishing of
a certain element of cohomology, the parameter $\tau$ is irrelevant in the new theory.  This means that the new theory
is really unique.

This new $\Z_2$-graded theory appears to be a candidate for the one
studied in \cite{BN}.  Electric-magnetic duality acts nontrivially
on the $\CP^1$ that parametrizes theories that come from four
dimensions.  But the new theory, being unique, must be invariant
under duality.  In particular, as duality  exchanges $G$ and
$G^\vee$, the new three-dimensional theory defined for $G$ is
equivalent to the same theory defined for $G^\vee$.

Starting with any point on the quadric $\mathcal Q$, corresponding
to one of the usual theories studied in (ordinary or quantum)
geometric Langlands, and making an infinitesimal perturbation away
from $\mathcal Q$, one lands on the same generic $F_\C$ orbit.  So
the same theory -- the one that is symmetrical between $G$ and
$G^\vee$ -- can be reached (after compactification to three
dimensions) by an infinitesimal perturbation of any of the theories
of four-dimensional origin.  The required perturbation  reduces the
$\Z$-grading to a $\Z_2$-grading.

\subsubsection{Vanishing Of A Certain Element Of Cohomology}\label{vanishing}

As explained in section \ref{complication}, the reason that the gauge coupling parameter $\tau$ is not completely
irrelevant in the twisted four-dimensional theories that lead to geometric Langlands is that the instanton number
$\nu=\int_M\P^{(4)}=\frac{1}{8\pi^2}\int_M\Tr\,F\wedge F$ is $Q$-invariant and not of the form $\{Q,\dots\}$ -- that is, it represents a nontrivial cohomology class of $Q$.   Adding a multiple of $\nu$ to the Lagrangian gives a non-trivial deformation
of the theory.

It turns out that when we perturb slightly away from the quadric, this cohomology class disappears.  As a result, the parameter
$\tau$ becomes irrelevant, completing the justification of the claim that after compactification to three dimensions on
a circle, there is precisely one new $\Z_2$-graded topological field theory that we can make.

\remark The fact that the cohomology class disappears under perturbation away from the quadric
 can be anticipated as follows. As shown in \cite{KW},
the deformation by the cohomology class $\nu$ is equivalent to the deformation
associated with a change in the linear combination $Q=uQ_++vQ_-$. We have already seen that once we move
away from $\mathcal Q$, the deformation by changing $Q$ becomes trivial, so the deformation by $\nu$ must also
become trivial.  Instead of relying on this sort of argument, we prefer to be more explicit.
\medskip

In general, for a cohomology class to disappear under an
infinitesimal perturbation, it must annihilate another cohomology
class whose $\Z$-grading differs  by $\pm 1$ (if the perturbation
preserves a $\Z$-grading, as in the case usually considered), or at
least one that has the opposite $\Z_2$ grading (if the perturbation
preserves only a $\Z_2$-grading, as in the case considered here). In
the four-dimensional topological field theories related to geometric
Langlands, there is no four-form valued cohomology class of $Q$ with
an odd grading that could possibly cancel $\int_M\P^{(4)}$ in the
cohomology.   However, once we compactify to three dimensions, there
is such a class.  Our construction on $M=M_3\times S^1$ made use of
a vector field $\VV$ that generates the rotation of $S^1$. There is
a natural $\VV$-invariant one-form $\d y$  on $S^1$ with $\int_{S^1
}\d y=1$. This enables us to consider the expression $\tilde
\nu=\int_M \P^{(3)}\wedge \d y$, which is a $Q$ cohomology class of
degree 1. If the four-dimensional $\Z$-graded theory is restricted
to four-manifolds of the form $M_3\times S^1$, then in addition to
the usual complex modulus corresponding to the cohomology class
$\nu$ (this modulus is tangent to the usual $\CP^1$ family), there
is an additional odd modulus corresponding to $\tilde \nu$.

But when one perturbs away from the quadric $\mathcal Q$ to a
$\Z_2$-graded theory, the cohomology classes $\nu$ and $\tilde\nu$
both disappear, as we will now argue.  Let  $Q$ be the topological
supersymmetry generator corresponding to a point in $\mathcal Q$ --
so $Q^2=0$ and $Q$ descends from four dimensions.  Pick a
one-parameter deformation $Q_\epsilon=Q+\epsilon Q'$, where $Q'$
corresponds to another point in $\PP(J)$ and $Q_\epsilon^2\not=0$.
After possibly replacing $Q'$ by a linear combination of $Q$ and
$Q'$, we can assume that $(Q')^2=0$ and
\begin{equation}\label{zelm}\{Q,Q'\}=\VV.\end{equation}

\def\CS{{\mathrm{CS}}}
Let $\CS(A)=\frac{1}{8\pi^2}\Tr\,\left(A\wedge dA+\frac{2}{3}A\wedge
A\wedge A\right)$ be the Chern-Simons three-form. Its periods are
not well-defined as real numbers, but rather take values in $\R/\Z$.
And let $\Theta=\int_M\CS(A)\wedge \d y$.   In defining $\Theta$, we
pick a point $y_0\in S^1$ and at that point, we pick a lift to $\R$
of $\int_{M_3\times{y_0}}\CS(A)$. Then we pick an $\R$-valued lift
of $f(y)=\int_{M_3\times y}\CS(A)$ so that this function is
continuous for $y>y_0$, and define $\Theta=\int_{S^1} \d y \,f(y)$.
Once we go all the way around the circle, $f(y)$ will jump by $\nu$,
the instanton number, so the definition of $\Theta$ depends on both
the choice of $y_0$ and the real lift chosen for
$\int_{M_3\times{y_0}}\CS(A)$. But the indeterminacy of $\Theta$ is
independent of $A$, and hence it makes sense to compute the
commutator $[\VV,\Theta]$, where $\VV$ acts on $A$ by generating the
rotation of the circle. Since $\int_{S^1}\d y (\d f/\d y)$ (which is
the change in $f$ in going around the circle) equals the instanton
number $\nu$, the commutator is
\begin{equation}\label{elf}[\VV,\Theta]=\nu.\end{equation}
(Physicists would usually describe this computation by saying that
$[\VV,A_i]=F_{y i }$,  where $A_i$ is a component of the connection
tangent to $M_3$ and $F_{yi}$ is a corresponding curvature
component. Using this, a formal evaluation of the commutator gives
(\ref{elf}).)

Another useful calculation gives
\begin{equation}\label{zext}[Q,\Theta]=\int_M \P^{(3)}\wedge \d y=\tilde\nu.\end{equation}
Again, the commutator makes sense because $\Theta$ is well-defined
modulo an additive constant. To compute this commutator, one needs
to know that $[Q,A]=\psi$, where $\psi$ is an adjoint-valued fermion
field such that $\P^{(3)}=\frac{1}{4\pi^2}\Tr\, F\wedge \psi$. The
formula (\ref{zext})  does not make $\tilde\nu=\int_M\P^{(3)}$
trivial in the cohomology of  $Q$, since $\Theta$ is not a
well-defined real-valued function.

However, now we find
$\{Q',\tilde\nu\}=\{Q',[Q,\Theta]\}=-\{Q,[Q',\Theta]\}+\{\VV,\Theta\}$,
where (\ref{zelm}) has been used along with the Jacobi identity.
Using also (\ref{elf}), we get
\begin{equation}\label{kozz}\{Q',\tilde\nu\}=\nu-\{Q,[Q',\Theta]\}.\end{equation}
Again, the commutator $[Q',\Theta]$ is well-defined despite the
uncertainty of $\Theta$ by a real constant (an explicit local
quantum field theory expression can be written for this commutator).
So when we pass to the cohomology of $Q$, the last term in
(\ref{kozz}) is trivial and this equation reduces to
$\{Q',\tilde\nu\}=\nu$.  This implies that when we perturb $Q$ to
$Q_\epsilon=Q+\epsilon Q'$, both $\tilde\nu$ and $\nu$ disappear
from the cohomology.

\remark \label{dorfox} Going back to four dimensions, we can select
an invariant polynomial $P_i$ of degree $d_i$ and perturb the
topological field theories related to geometric Langlands by the
$Q$-invariant interaction $\int_M \P_i^{(4)}$. This perturbation has
degree $2d_i-4$, so, for $d_i>2$, it gives a  $\Z_{2d_i-4}$-graded
theory. If we include a linear combination of such perturbations
with all possible values of $i$, we will get a family of
four-dimensional topological field theories that (for most simple
Lie groups $G$) are generically only $\Z_2$-graded.  These theories
have similar behavior under electric-magnetic duality to the
$\Z$-graded theories that are usually considered in geometric
Langlands. It is not clear to the author whether they contain any
essentially new information.

\subsection{Uniqueness Of The Solution Of Nahm's
Equation}\label{realdetails}

An important point in section \ref{subtlety} was that, with the
appropriate boundary conditions at the two ends, the solution of
Nahm's equations on the half-open interval $(0,L]$ is unique.  The
boundary condition   for $y\to 0$ was described in eqn.
(\ref{zonko}): $\vec X$ should have a regular pole at $y=0$, the
singular part being
\begin{equation}\label{ploonko}\vec X=\frac{\vec t}{y},\end{equation}
where $\vec t$ are the images of the $\frak{su}(2)$ generators under a principal embedding $\vartheta:\frak{su}(2)\to
\frak g^\vee$.

The boundary condition at $y=L$ was not explained in section
\ref{gaugetheory}, but as we will explain, its effect is that the
solutions of Nahm's equations on $(0,L]$ with the conditions we will
want at $y=L$ are tautologically the same as the solutions of Nahm's
equations on the open half-line $(0,\infty)$ with a requirement that
$\vec X\to 0$ at infinity.

Kronheimer \cite{Kr} investigated Nahm's equations on the open
half-line with these conditions\footnote{Kronheimer also considered
a generalization of the condition $\vec X\to 0$ at infinity, the
requirement being instead that $\vec X$ is conjugate at infinity to
a specified triple of elements of $\vec t^\vee$.  It is possible to
modify our boundary conditions on both the $\A$-model and $\B$-model
side so as to arrive at this generalization.  The necessary facts
are mostly presented in \cite{GW1}.  But we will omit this
generalization here.} (including the regular Nahm pole at $y=0$)
 and showed that the solution
is unique.  So once we explain how our problem on the half-open interval $(0,L]$ is related to Kronheimer's problem
on the half-line $(0,\infty)$, the uniqueness claimed in section \ref{subtlety} will follow.

Actually, Kronheimer considered a more general problem in which
$\vartheta:\frak{su}(2)\to \frak g^\vee$ is taken to be an arbitrary
homomorphism, not necessarily related to a principal embedding.  We
will need to know about the opposite case $\vartheta=0$. For this
choice, there is no pole at $y=0$, so one is studying solutions of
Nahm's equations on the closed half-line $[0,\infty)$.  In this
case, the moduli space of solutions of Nahm's equations turns out to
be a hyper-Kahler manifold $\mathcal X(G^\vee)$ that in any of its
complex structures is equivalent to the nilpotent cone in the
complex Lie algebra $\frak g^\vee_\C$.  The moduli space $\mathcal
X(G^\vee)$ has $G^\vee$ symmetry for an easily understood reason: if
$\vartheta=0$, then the group $G^\vee$ acts on the solutions of
Nahm's equations in the obvious fashion $\vec X\to g\vec X g^{-1}$.
(For $\vartheta\not=0$, the group that acts is the subgroup of
$G^\vee$ that commutes with the image of $\vartheta$.) The
hyper-Kahler moment map for the $G^\vee$ action on $\mathcal
X(G^\vee)$ turns out to be $\vec \mu =\vec X(0)$.  All this has the
following  trivial generalization. If we solve Nahm's equations on
the half-line $[L,\infty)$ (rather than $[0,\infty)$), we get an
isomorphic hyper-Kahler manifold,  the moment map for the $G^\vee$
action now being
\begin{equation}\label{getz} \vec\mu=\vec X(L).\end{equation}

Here we will only require the extreme cases that $\vartheta$ is
either 0 or a principal embedding. The general result \cite{Kr},
however, for any $\vartheta$, is that the moduli space of solutions
of Nahm's equations turns out to be, as a complex manifold in any of
its complex structures, the Slodowy slice transverse to the
nilpotent element $t_1+it_2$ of $\frak g^\vee_\C$.

Now we need to describe the boundary conditions at $y=L$ in the
construction of section \ref{subtlety}. The relevant notion of a
boundary condition is more extended than one may be accustomed to in
the world of partial differential equations, for example.  A
boundary condition in a quantum field theory defined on
$d$-manifolds is a choice of how to extend the definition to
$d$-manifolds with boundary in such a way that all the usual axioms
of local quantum field theory are preserved.  This notion allows us
to include on the boundary a $d-1$-dimensional quantum field theory.
It is only interesting to do that, however, if the $d-1$-dimensional
boundary theory is coupled in some way to the ``bulk'' theory.

One might think that this notion of a boundary condition is too broad.  However, it is shown in \cite{GW1,GW2} that this
extended notion of a boundary condition is unavoidable if one wishes electric-magnetic duality to act on boundary conditions,
since the dual of a more conventional boundary condition can very well be a boundary condition in this extended sense.
For example, as shown in \cite{GW2}, the dual of Dirichlet boundary conditions in $G$-gauge theory is a boundary
condition in $G^\vee$ gauge theory that involves coupling of the $G^\vee$ gauge fields to a very special superconformal
field theory $T(G^\vee)$ that is supported on the boundary.  For our purposes,  $T(G^\vee)$ has the following important
properties.  It has $G^\vee\times G$ as a group of global symmetries.   The Higgs branch of vacua of $T(G^\vee)$ turns out
to be the Kronheimer manifold $\mathcal X(G^\vee)$, and the Coulomb branch of vacua is the dual Kronheimer manifold
$\mathcal X(G)$.

As is explained in \cite{GW1}, for a boundary condition in $G^\vee$
gauge theory that is obtained by coupling to a boundary theory with
$G^\vee$ symmetry, the appropriate boundary condition in Nahm's
equations is to set $\vec X$ equal on the boundary to $\vec \mu$,
the moment map for the action of $G^\vee$ on the Higgs branch:
\begin{equation}\label{olp}\vec X(L)=\vec\mu.\end{equation}
This equation looks just like (\ref{getz}), even though the two
equations have a completely different meaning.  In (\ref{olp}),
$\vec X$ is a solution of Nahm's equations on the interval $(0,L]$
where the quantum field theory is defined. In eqn. (\ref{getz}),
$\vec X$ is a solution of Nahm's equations on the half-line
$[L,\infty)$.  It defines a point in the Higgs branch of the
boundary theory.  Nevertheless, if we simply combine the two
equations, we see that, even though their interpretations are
completely different, the solution of Nahm's equations on $(0,L]$
agrees at $y=L$ with the solution of Nahm's equations on
$[L,\infty)$.  Hence, they fit together to a single solution of
Nahm's equations on the open half-line $(0,\infty)$.  Nahm's
equations ensure that this solution is smooth near $y=L$. It has the
singular behavior (\ref{ploonko}) near $y=0$, and vanishes for
$y\to\infty$ since this is a characteristic of the moduli space
$\mathcal X(G)$. According to the first result of Kronheimer
mentioned at the beginning of this subsection, Nahm's equations have
a unique solution (namely $\vec X=\vec t/y$) obeying these
conditions. This is the uniqueness asserted in section
\ref{subtlety}.

  The examples that we
have described here of the role of Nahm's equations in duality of
boundary conditions in $\NN=4$ super Yang-Mills theory are really
only the tip of the iceberg. Much more can be found in
\cite{GW1,GW2}. The full story involves, among other things, the
more general moduli spaces defined by Kronheimer for an arbitrary
$\vartheta:\frak {su}(2)\to \frak{g}$.

\subsection{More On The Dual Of Dirichlet Boundary Conditions}\label{genker}

In section \ref{realdetails}, we exploited in a rather technical way
the special properties of the dual of Dirichlet boundary conditions.
We perhaps should not leave the subject without explaining that the
dual of Dirichlet boundary conditions actually plays a rather basic
role in the geometric Langlands correspondence.

\def\Br{{\mathcal{B}}}
We start by explaining intuitively why the dual of Dirichlet
boundary conditions should be important.  In geometric Langlands,
one considers the $\B$-model of $\NN=4$ super Yang-Mills theory,
compactified on a Riemann surface $C$, for gauge group $G^\vee$. One
compares it to the $\A$-model of $G$ on the same Riemann surface.
The most basic branes in the $\B$-model are branes associated with a
homomomorphism $\chi:\pi_1(C)\to G^\vee_\C$.  One would like to
understand their duals in the $\A$-model.

Let us start with the case that $\chi$ is trivial.  Let $\Br$ be the
corresponding $\B$-brane.  We could modify $\Br$ by introducing a
Nahm pole, but let us not do so.

Then the brane $\Br$ is simply the one that is defined by Dirichlet
boundary conditions for the complexified gauge field $\AA=A+i\phi$
(extended to all other fields to preserve the topological
supersymmetry of the $\B$-model).  After all, Dirichlet boundary
conditions say that $\AA$ should be trivialized on the boundary, so
that the boundary data correspond to a trivial flat connection
representing the  trivial homomorphism from $\pi_1(C)$ to
$G^\vee_\C$.

Dirichlet boundary conditions can be considered without any
compactification, as indeed was done in \cite{GW1,GW2}.  Thus the
brane $\Br$ associated to the trivial flat connection without a Nahm
pole has a universal meaning, independent of any choice of Riemann
surface $C$.  (This is also true for the analogous problem with a
specified Nahm pole.)

Let $\Br^*$ be the $\A$-brane that is dual to $\Br$.  Then $\Br^*$,
like $\Br$, can be defined universally without any choice of
compactification.  As explained in \cite{GW2} and as we already
stated in section \ref{realdetails}, $\Br^*$ is defined by coupling
$G$ gauge theory to a three-dimensional superconformal field theory
$T(G)$ that has $G\times G^\vee$ global symmetry.\footnote{For
three-dimensional superconformal field theories with the relevant
amount of supersymmetry, there is a notion of mirror symmetry
\cite{IS}, somewhat analogous to the more familiar mirror symmetry
in two dimensions.  The mirror of $T(G)$, in this sense, is
$T(G^\vee)$. Indeed, $T(SU(2))$ was one of the fundamental examples
considered in \cite{IS}.}  One uses the $G$ symmetry of $T(G)$ to
couple it to $G$ gauge fields in bulk. This leaves a $G^\vee$ global
symmetry, matching the fact that $G^\vee$ is the automorphism group
of Dirichlet boundary conditions (or of the trivial homomorphism
$\pi_1(C)\to G^\vee_\C$) in $G^\vee$ gauge theory.

The duality between $\Br$ and $\Br^*$ holds before or after
compactification on a Riemann surface $C$.  However, after
compactification, we can consider a twisted version of the picture
in which we twist using the automorphism group $G^\vee_\C$, which
$\Br$ and $\Br^*$ have in common. On the $\B$-model side, the
twisted version of the picture simply involves a choice of
homomorphism $\chi:\pi_1(C)\to G^\vee_\C$.  To each choice of
$\chi$, one defines a $\B$-brane $\Br(\chi)$ that is locally
isomorphic to $\Br$, but globally is obtained from $\Br$ by twisting
by the homomorphism $\chi$ from $\pi_1(C)$ to the automorphism group
$G^\vee_\C$ of $\Br$. (The statement that $\Br(\chi)$ is ``locally''
isomorphic to $\Br$ means that they are isomorphic locally along
$C$.)  Let us denote as $\Br^*(\chi)$ the dual of $\Br(\chi)$. Then
$\Br^*(\chi)$ is obtained from $\Br^*$ exactly as $\Br(\chi)$ was
obtained from $\Br$: by twisting via a homomorphism from $\pi_1(C)$
to the automorphism group.  This makes sense since $\Br$ and $\Br^*$
have the same automorphism group $G^\vee$.

So the dual of any $\Br(\chi)$ can be constructed if one understands
the three-dimensional superconformal field theory $T(G)$ that is the
main ingredient in describing the dual of Dirichlet boundary
conditions.  Thus, a knowledge of $T(G)$ gives the same sort of
results that one would expect mathematically from a description of
the universal kernel of geometric Langlands.  This universal kernel
is supposed to be a brane in the product theory $\A(G)\times
\B(G^\vee)$ that has certain universal properties.  In fact, $T(G)$
can be used to construct the appropriate brane. This can be done
prior to compactification, and thus independent of any choice of
$C$.

The relevant construction is quite simple and was described in
section 4 of \cite{GW2}.  One divides $\R^4$ into two half-spaces
separated by a copy of $\R^3$, supported at, say, $y=0$, where $y$
is one of the Euclidean coordinates of $\R^4$. For $y<0$, one places
$\NN=4$ super Yang-Mills theory with gauge group $G$; for $y>0$, one
places the same theory with gauge group $G^\vee$. At $y=0$, one
places the theory $T(G)$. Using its $G\times G^\vee$ global
symmetries, it can be coupled to $G$ gauge theory on the left and
$G^\vee $ gauge theory on the right. Moreover, the coupling can be
chosen so that the whole construction is supersymmetric -- to be
more precise, invariant under a subgroup $OSp(4|4)$ of the symmetry
supergroup $PSU(2,2|4)$ of $\NN=4$ super Yang-Mills theory.  One can
pick a fermionic generator of $OSp(4|4)$ that for $y<0$ generates
the topological supersymmetry of the $\A$-model of $G$, and for
$y>0$ generates the corresponding symmetry of the $\B$-model of
$G^\vee$.

To get closer to the usual mathematical point of view, we can
``fold'' $\R^4$ along the hypersurface $y=0$, so that the $G$ and
$G^\vee $ gauge groups now both are supported at $y<0$ and there is
nothing for $y>0$.  In this description, then, the theory $T(G)$
provides a boundary condition in the product of $G$ and $G^\vee$
gauge theories. After topological twisting, this boundary condition
corresponds to a brane $\tilde\Br$ in the product of the $\A$-model
of $G$ and the $\B$-model of $G^\vee$.  Like the branes $\Br$ and
$\Br^*$ discussed above, $\tilde\Br$ can be defined in a universal
way without any compactification.  This indeed was the viewpoint in
\cite{GW2}, where properties were discussed that correspond to the
desired universal properties in geometric Langlands.

\bigskip
{\it Acknowledgments~} The idea of connecting the Langlands
correspondence with gauge theory was first proposed in the 1970's by
M. F. Atiyah (motivated by the observation that the
Goddard-Nuyts-Olive dual group is the same as the Langlands dual
group).  I am grateful to him for introducing me to those ideas at
that time.  I also would like to thank S. Cherkis as well as the referee for a close
reading of the manuscript and careful comments, E. Frenkel for some
helpful questions, and D. Gaiotto for collaboration on
electric-magnetic duality of boundary conditions.

\def\H{{\mathcal H}}
\def\cal{\mathcal}
\bibliographystyle{amsalpha}

\end{document}